\documentclass[journal]{IEEEtran}
\usepackage{cite}
\usepackage{amsmath,amssymb,amsfonts}
\usepackage{graphicx,setspace}
\usepackage{textcomp}
\usepackage{titlesec}
\usepackage{amssymb}
\newcommand{\cmark}{\checkmark}
\newcommand{\xmark}{\text{\small $\times$}}

\titlespacing*{\section}{0pt}{1ex}{0.5ex}   
\titlespacing*{\subsection}{0pt}{1ex}{0.5ex}
\titlespacing*{\subsubsection}{0pt}{1ex}{0.5ex}
\usepackage[table,xcdraw]{xcolor} 
\usepackage{dsfont}
\usepackage{algorithm,algpseudocode}
\newcommand\CONDITION[2]%
  {\begin{tabular}[t]{@{}l@{}l@{}}
     #1&#2
   \end{tabular}%
  }
  \usepackage{multirow} 
\usepackage{array}    
\algdef{SE}[WHILE]{While}{EndWhile}[1]%
  {\algorithmicwhile\ \CONDITION{#1}{\ \algorithmicdo}}%
  {\algorithmicend\ \algorithmicwhile}
\algdef{SE}[FOR]{For}{EndFor}[1]%
  {\algorithmicfor\ \CONDITION{#1}{\ \algorithmicdo}}%
  {\algorithmicend\ \algorithmicfor}
\algdef{S}[FOR]{ForAll}[1]%
  {\algorithmicforall\ \CONDITION{#1}{\ \algorithmicdo}}
\algdef{SE}[REPEAT]{Repeat}{Until}{\algorithmicrepeat}[1]%
  {\algorithmicuntil\ \CONDITION{#1}{}}
\algdef{SE}[IF]{If}{EndIf}[1]%
  {\algorithmicif\ \CONDITION{#1}{\ \algorithmicthen}}%
  {\algorithmicend\ \algorithmicif}%
\algdef{C}[IF]{IF}{ElsIf}[1]%
  {\algorithmicelse\ \algorithmicif\ \CONDITION{#1}{\ \algorithmicthen}}
\usepackage{textcomp}
\usepackage{nccmath}
\usepackage{array}
\usepackage{tabularx}
\usepackage{booktabs}
\usepackage{mathtools}
\usepackage{comment}
\usepackage{amsthm}
\usepackage{setspace} 
\IEEEoverridecommandlockouts
\usepackage{url}
\usepackage{siunitx}
\usepackage{pgfplots}
\usepackage{epstopdf}
\pgfplotsset{compat=1.9}
\usepackage{pgfplots,capt-of}

\usepackage{scalefnt}
\usepackage{bigints}
\usepackage{booktabs} 
\usepackage{longtable} 
\usetikzlibrary {arrows.meta}
\usepackage{soul}   

\sethlcolor{blue!20}
\usepackage{epstopdf}

\usepackage{subfigure}
\usepackage[tableposition=top]{caption}
\DeclareCaptionFont{allcaps}{\MakeUppercase}
\newtheorem{theorem}{Theorem}
\theoremstyle{theorem}
\theoremstyle{definition}
\usepackage[dvipsnames]{xcolor}
\usetikzlibrary{spy}
\usepackage{mathrsfs}
\newcommand{\startmycolor}{%
    \color{black}
}

\newcommand{\stopmycolor}{%
    \color{black}
}

\newtheorem{proposition}{Proposition}
\newtheorem{lemma}{Lemma}

\newtheorem{corollary}{Corollary}

\usepackage{mdframed}
\DeclareCaptionFormat{colored}{%
  \begin{mdframed}[backgroundcolor=blue!10,linecolor=blue!40,linewidth=0pt,innertopmargin=2pt,innerbottommargin=2pt,innerleftmargin=4pt,innerrightmargin=4pt]%
    #1#2#3%
  \end{mdframed}%
}

\allowdisplaybreaks
\begin{document}
\title{Merits of Collocating Uptilted Antennas on Terrestrial Base Stations for Aerial Coverage}

\author{%
Nesrine~Cherif,~\IEEEmembership{Member, IEEE}, and
Qurrat-Ul-Ain Nadeem,~\IEEEmembership{Senior Member, IEEE}

\thanks{N. Cherif is with the Engineering Division, New York University (NYU) Abu Dhabi,  PO Box 129188, Abu Dhabi, UAE (e-mail: nc3654@nyu.edu).}%
\thanks{Q.-U.-A. Nadeem is with the Engineering Division, New York University Abu Dhabi (NYUAD), PO Box 129188, Abu Dhabi, UAE, and also with the  Tandon School of Engineering, New York University, Brooklyn, NY 11201, USA. She is also with NYUAD Wireless Center, NYUAD Research Institute, PO Box 129188, Abu Dhabi, UAE (e-mail: qurrat.nadeem@nyu.edu).}%
\thanks{The work of Q.-U.-A. Nadeem was supported by Tamkeen under the NYUAD Research Institute grant CG017.}
}

\maketitle
\linespread{1.1}
\begin{abstract}
The integration of unmanned aerial vehicles (UAVs) into future wireless networks introduces challenges in ensuring reliable aerial coverage, particularly for UAV-based user equipment (UAV-UE). Traditional terrestrial base stations (BSs) with downtilted antennas, referred to as dt-BSs, are designed primarily for ground users. They provide inefficient connectivity to UAV-UE due to weak sidelobe gains, and susceptibility to ground reflection (GR) interference as well as sidelobe interference from non-serving dt-BSs.  To address this, we propose a hybrid cellular architecture that collocates uptilted antennas on a portion of existing dt-BS towers, referred to as ut-BSs. These support UAV-UE exclusively, minimizing impact on ground users. We consider realistic antenna radiation patterns and probabilistic line-of-sight (LoS)/non-LoS (NLoS) links between ut-BSs and UAV-UE. Using stochastic geometry, we analyze the UAV-UE coverage probability, accounting for interference from both dt-BSs and non-serving ut-BSs, along with GR-induced interference. 
Results show significant improvements in UAV-UE coverage with only a small density of ut-BSs. GR interference is notable at lower altitudes but diminishes beyond 500 meters. This study highlights the potential of our hybrid architecture to enhance the reliability and scalability of UAV communications in large-scale deployments. It also offers actionable insights for telecom operators considering the collocation of uptilted antennas within existing infrastructure to better support aerial connectivity.

\end{abstract}

\section{Introduction}

\IEEEPARstart{W}{ith} the rapid development and declining costs of unmanned aerial vehicle (UAV) technology, we can  anticipate widespread adoption of UAVs for various use-cases, such as last-mile cargo delivery, surveillance, smart agriculture, search-and-rescue operations, and more. Yet, guaranteeing ubiquitous and reliable cellular connectivity for these  emerging aerial user equipment (UE), known as UAV-UE, poses a considerable challenge. UAV-UE can establish line-of-sight (LoS) transmission links with terrestrial base stations (BSs) by dynamically adjusting their three-dimensional positions. However, they can only be served through the sidelobes of the BS that is serving ground UE (GUE), and they are prone to high interference over LoS links from non-serving BSs. As a result, the distinctive operational traits and connectivity needs of UAV-UE require an effective re-use of conventional cellular network infrastructure, which has  historically been optimized for GUE. 

\subsection{Motivation and Related Work}
Terrestrial cellular networks are primarily designed for GUE, with downtilted antennas at the BSs that enhance ground coverage and limit inter-cell interference.  When terrestrial networks are reused to serve UAV-UE, downtilted antenna radiation patterns provide limited signal gain toward the sky while exposing UAV-UE to strong interference from neighboring cells. As a result, a straightforward reuse of existing cellular infrastructure to serve UAV-UE is often inadequate, and tailored solutions are needed to account for the distinct propagation conditions experienced by UAV-UE \cite{mei2021aerial}. This challenge has been highlighted in the 3GPP TR 36.777 \cite{3gpp777}, which reports that the sidelobe gain of downtilted antennas is limited at higher elevations, while the interference levels experienced by aerial users are typically high.

In response, several works have proposed adapting terrestrial BS configurations to better support UAV-UE. The authors in \cite{chen32023} suggested converting a subset of downtilted BSs into uptilted BSs to enhance UAV-UE service. While effective for aerial coverage, this approach inherently introduces a performance trade-off, as \textcolor{black}{redirecting transmit antennas} away from the ground will degrade the quality of service at GUE. Relatedly, \cite{karimi2024optimizing} considers a network-wide optimization framework in which the BS antenna tilts and transmit powers are jointly optimized via quantization-theoretic tools to balance the signal quality between GUE and UAV-UE. Importantly, this model does not rely on explicit time-/frequency-division switching between user types, rather, the same cellular infrastructure is concurrently optimized to serve a mixed aerial and ground user population. Nevertheless, the fundamental trade-off persists because the same BS resources, i.e., radiation pattern and power, must be shared between terrestrial and aerial coverage objectives. More recently, \cite{qin2025stochastic} investigated cell-free massive MIMO (CF-mMIMO) systems with aerial users, where a fraction of downtilted BSs is converted into uptilted ones. Closed-form signal-to-interference-and-noise ratio (SINR) and rate coverage expressions were derived, but the considered architecture, based on uniform linear array (ULA) of antennas, was shown to offer limited improvements due to dominant interference from non-serving BSs.  Deploying additional terrestrial infrastructure (e.g., new BS towers with uptilted arrays dedicated to UAV-UE) can be considered, but is typically not cost-effective from an operator’s perspective.

Another important limitation of much of the existing literature on cellular connectivity for UAV-UE is the treatment of interference. In particular, the impact of ground reflection (GR) of the mainlobe of the downtilted BS array's radiation pattern at the UAV-UE has often been neglected, despite its potential to induce strong interference at elevated UAV-UE heights. Typically, only the interference originating from the sidelobes of non-serving BSs is considered and modeled at the UAV-UE, which is an underestimation of the overall interference experienced by the UAV-UE. To model GR more accurately, \cite{chowdhury2021ensuring} modified the height-dependent air-to-ground path-loss model of \cite{goddemeier2012role} to include angle-dependent antenna gains, thereby capturing the coupling between BS downtilt, UAV altitude, and GR of the mainlobe. Note that before UAVs emerged as a distinct user class, GR from downtilted mainlobes at elevated heights was largely irrelevant, since users were confined near ground level. With the integration of UAVs into existing cellular networks, however, GR becomes a non-negligible component of the interference field and must be considered in the design and performance evaluation of connectivity solutions. The results in \cite{chowdhury2021ensuring} have shown that GR-induced interference can severely degrade the signal-to-interference (SIR) at UAV-UE, motivating the use of enhanced inter-cell interference coordination (eICIC) schemes to mitigate this effect. Unfortunately, such interference coordination significantly reduces the overall spectral efficiency of the network.

Overlay architectures that equip existing BSs with additional uptilted antenna arrays have  been investigated in \cite{chowdhury2021ensuring,singh2020spectrum,bhuyan2022advances}. These works typically assume ULA-based uptilted antennas and focus on SIR analysis at UAV-UE, demonstrating only modest performance gains. Moreover, they often adopt a full overlay strategy, where uptilted antennas are collocated at all downtilted BSs (dt-BSs), which limits design flexibility and may introduce unnecessary interference and cost. In contrast, there remains a need for the design and analysis of overlay architectures that (i) explicitly account for all sources of interference at the UAV-UE, (ii) leverage  directional uptilted antennas  to better isolate aerial links from terrestrial ones, and (iii) allow partial collocation of uptilted antenna arrays to balance performance and deployment cost.

Motivated by these observations and the limitations highlighted above, this work proposes a hybrid overlaid architecture that augments a subset of existing dt-BSs with additional uptilted antenna arrays dedicated to serving UAV-UE, while explicitly incorporating ground-reflected mainlobe and direct sidelobe interference from dt-BSs into the analysis.

To position our work within the existing literature on cellular connectivity solutions for UAV-UE, we provide a comparative summary of representative studies across several key dimensions. Specifically, Table~\ref{tab:related} indicates whether each work proposes an adapted architecture for UAV-UE connectivity, incorporates uptilted antennas, supports partial overlay, employs highly directional uniform planar array (UPA) for uptilted antennas, explicitly accounts for interference arising from GR of the downtilted radiation pattern, and adopts a stochastic geometry (SG) framework for network-level performance analysis.

\begin{table}[t!] \centering \caption{\textcolor{black}{Features considered in representative works on cellular connectivity for UAV-UE}} \label{tab:related} \footnotesize \setlength{\tabcolsep}{2.5pt} 
 {
\begin{tabular}{lcccccc} \hline Work & \shortstack{Adapted\\architecture} & \shortstack{Uptilted\\antennas} & \shortstack{Partial\\overlay} & \shortstack{UPA ut-\\ antennas} & \shortstack{GR-\\aware} & \shortstack{SG-based\\analysis} \\ \hline 
3GPP TR \cite{3gpp777} & \xmark & \xmark & \xmark & \xmark & \xmark & \xmark \\ Chen \cite{chen32023} & \cmark & \cmark & \cmark & \xmark & \xmark & \cmark \\ 
Karimi \cite{karimi2024optimizing} & \cmark & \cmark & \xmark & \xmark & \xmark & \xmark \\ 
Qin \cite{qin2025stochastic} & \cmark & \cmark & \cmark & \xmark & \xmark & \cmark \\ 
Chowdhury \cite{chowdhury2021ensuring} & \cmark & \cmark & \xmark & \xmark & \cmark & \xmark \\ 
Singh \cite{singh2020spectrum} & \cmark & \cmark & \xmark & \xmark & \cmark & \xmark \\ 
Bhuyan \cite{bhuyan2022advances} & \cmark & \cmark & \xmark & \xmark & \cmark & \xmark \\ \textbf{This work} & \cmark & \cmark & \cmark & \cmark & \cmark & \cmark \\ \hline \end{tabular} 
}
\vspace{10pt}
\end{table}

\subsection{Contributions}
In this context, the main contributions of this paper are summarized as follows:
\begin{itemize}

\item \textbf{Hybrid overlaid cellular architecture with partial collocation:}
We propose a novel cellular architecture for enhancing aerial coverage in which a portion of existing downtilted BSs (dt-BSs) is overlaid with additional uptilted antenna arrays dedicated to serving UAV-UE. The resulting upgraded BSs, equipped with both downtilted and uptilted antennas, are referred to as ut-BSs. Transmissions from ut-BSs share the same time-frequency resources as those from dt-BSs; however, the mainlobes of the additional uptilted arrays are steered toward the sky, thereby providing reliable connectivity to UAV-UE while preserving the coverage of GUE. The proposed solution requires only the installation of additional antenna arrays and RF chains on existing towers, and does not require new radio tower deployment. Unlike \cite{chowdhury2021ensuring,singh2020spectrum,bhuyan2022advances},  which assume uptilted ULAs  and full collocation at all BSs, our architecture employs uptilted   UPAs and a partial collocation strategy, offering a more flexible and efficient design space.

\item \textbf{Stochastic geometry-based analytical framework with GR-aware interference modeling:}
We develop a comprehensive analytical framework based on tools from stochastic geometry to characterize the coverage probability of a typical UAV-UE under the proposed hybrid architecture. The analysis incorporates both LoS and non-LoS (NLoS) links between ut-BSs and the UAV-UE. We explicitly account for three major sources of interference at the UAV-UE: (i) interference from the sidelobes of dt-BSs, (ii) interference arising from GR of the mainlobes of dt-BSs, and (iii) interference from non-serving ut-BSs. For each component, we derive the corresponding Laplace transform, which is then used to obtain analytical expressions for the UAV-UE coverage probability. A key contribution of this work is the comprehensive analytical interference characterization at aerial users in such an overlaid architecture.

\item \textbf{Performance evaluation and design guidelines for UAV-UE coverage:}
We present extensive simulation results to validate the analytical framework and to quantify the coverage probability of UAV-UE under the proposed architecture. First, we compare the performance of the hybrid overlaid design, where a subset of dt-BSs is additionally designated as ut-BSs, against baseline strategies such as reusing only dt-BSs or fully converting a portion of dt-BSs into ut-BSs. The results demonstrate significant coverage gains for UAV-UE under the proposed design with negligible loss in GUE performance. Then, we investigate the impact of dt-BS parameters, including the downtilt angle and ground type, on UAV-UE performance.  Finally, we formulate and solve an optimization problem to  provide concrete recommendations on the UAV-UE height, overlay percentage (i.e., fraction of dt-BSs that should be additionally equipped with uptilted arrays), and transmit powers that maximize UAV-UE coverage while ensuring that the performance degradation experienced by GUE due to ut-BS interference remains within acceptable limits.

\end{itemize}

\stopmycolor
The remainder of the paper is organized as follows. In Section II, we detail the system model,  the channel models, and the radiation patterns of the dt-BSs and ut-BSs. Section III focuses on the coverage probability analysis of the UAV-UE, which is then used to formulate an optimization problem to design the ut-BS parameters in Sec. IV. An extensive set of numerical results is provided in Section V. Finally, we conclude this work in Section VI.
\section{System model}

In this section, we outline the network architecture, the antenna radiation patterns, and the channel models.

\subsection{Network Setup}

We assume an area where the tier of dt-BSs, $\Phi_{\rm DT}$, is  distributed as a Poisson point process (PPP) with a density $\lambda_{\rm dt}$ per km$^2$. The dt-BSs are assumed to have equal transmit powers  $P_{\rm dt}$, and are equipped with $N_t$ vertically arranged  directional antennas placed at a height of $h_{\rm dt}$ from the ground \cite{chowdhury2021ensuring}.  We assume these antennas to have an omnidirectional radiation pattern in the azimuth plane and a controllable radiation pattern in the elevation plane that is defined in the next section.

To provide seamless and ubiquitous cellular coverage for aerial users at elevated heights, we overlay some of the dt-BSs with directional uptilted antenna arrays. These dt-BSs are additionally referred to as ut-BSs in the rest of the paper.  The tier of the ut-BSs, i.e., set of BSs with an uptilted antenna array installed in addition to the existing downtilted array, is assumed to be a PPP, $\Phi_{\rm UT}$,  with density  $\lambda_{\rm ut}$ per km$^2$. It is worth emphasizing that a subset of the dt-BSs is chosen to simultaneously serve as ut-BSs. This way we do not need to deploy dedicated towers to provide coverage to aerial users, but rather re-purpose the existing towers with co-existing uptilted and downtilted antenna arrays. To limit the electromagnetic interference (EMI) from the newly installed uptilted antennas, we assume that this array will be at an altitude of $h_{\rm ut}$ with respect to the ground that is greater than $h_{\rm dt}$\footnote{To avoid mutual coupling between the uptilted and downtilted antennas, we assume a vertical separation multiple wavelengths \cite{itu2011}.}. We assume that the uptilted antennas are arranged in a uniform planar array (UPA), i.e., in the x-y plane, \cite{venugopal2016device,yi2019unified,guo2021coverage}. Finally, we assume that the transmission power of each ut-BS is equal to  $P_{\rm ut}$.

\begin{figure}[t!]
	\centering
	\includegraphics[scale=0.3]{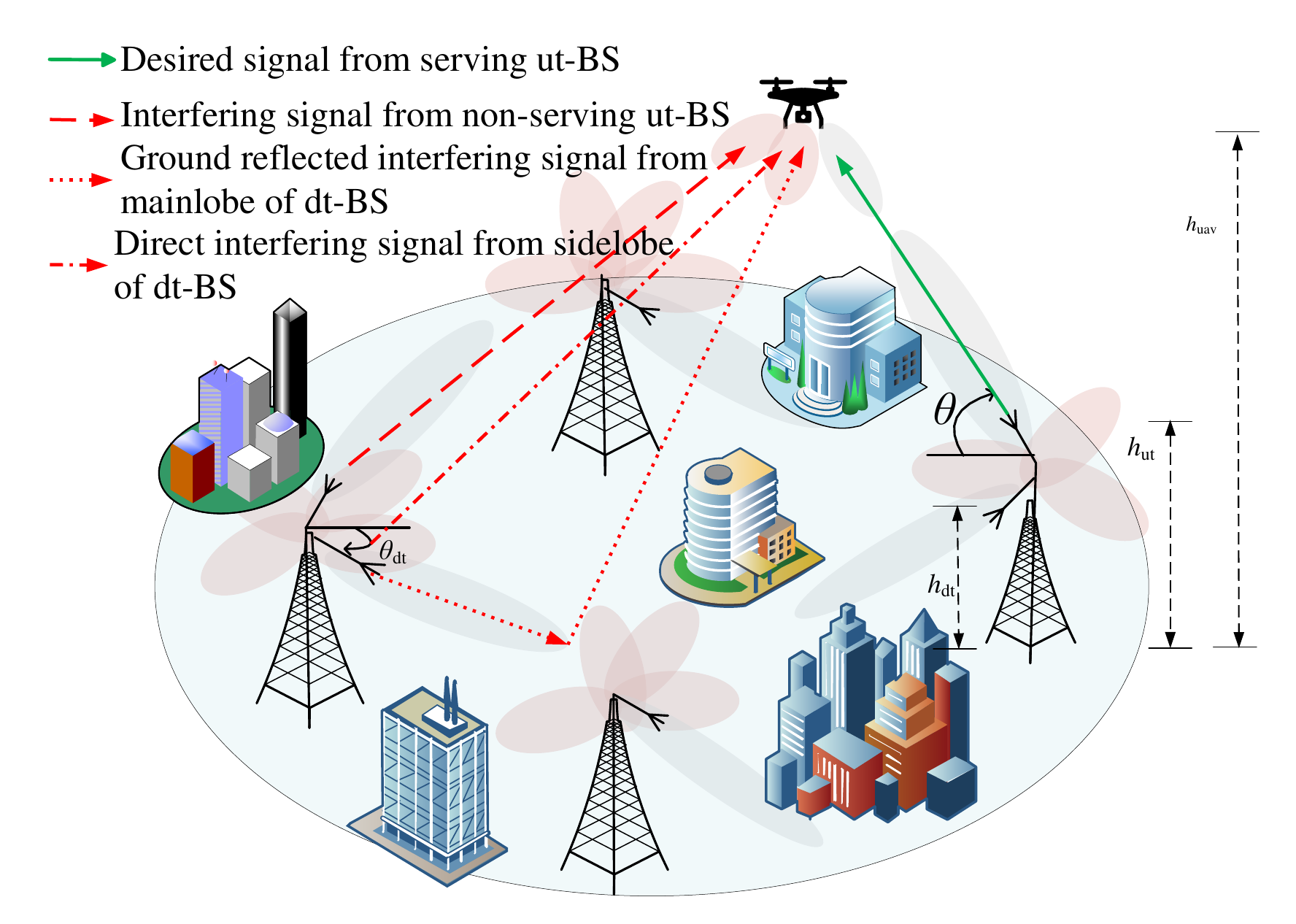}
	\caption{Network architecture for cellular-connected UAV-UE with inter-cell interference. }
    \vspace{10pt}
	\label{fig:sys}
\end{figure}
It has been extensively investigated in the literature that serving aerial users with existing terrestrial networks, which have BSs with antenna radiation patterns downtilted towards the GUE, results in poor performance in terms of reliability and coverage \cite{cherifdownlink}. Thus, we consider that UAV-UE will only be served by the ut-BSs that have dedicated antenna arrays installed for the purpose of providing strong aerial coverage. The uptilted and downtilted antennas are assumed to be operating simultaneously over all resource blocks unless stated otherwise throughout the paper. Finally, we follow the \textcolor{black}{strongest average received power} policy to associate the aerial users with ut-BSs. In the rest of the work and without loss of generality, we will focus on one typical UAV-UE located at the origin with coordinates $(0,0,h_{\rm uav})$, the network of dt-BSs, and the overlaid network of ut-BSs. 

\subsection{Antenna Radiation Pattern of dt-BSs}
The dt-BS vertical radiation pattern is designed primarily to support the GUE cellular coverage. However, it can cause some interference at the UAV-UE, which mainly depends on its 3dB beamwidth and the downtilt angle. In order to characterize it, we measure its value at an elevation angle corresponding to the UAV-UE. In the following, we start by characterizing the downtilted antenna radiation pattern,  and then we measure the amount of interference it causes at the typical UAV-UE.

The dt-BSs are equipped with ULAs of $N_t$ equally spaced antenna elements with an inter-element spacing of half-wavelength. The antennas are omnidirectional in the azimuth plane. \textcolor{black}{The antenna element gain in the elevation in dB  is expressed using \cite[Table A.2.1.1-2 ]{3gpp36814} as follows}
\begin{equation}
    \label{eq:antennavertical}
    G_e^{\rm dt}(\theta)=G_{\max}-\text{min} \left[12\left(\frac{\theta-\theta_{\rm dt}}{\theta_{\rm 3dB}^{\rm dt}}\right)^2\text{, }G_{\rm m}\right],
\end{equation}
where $G_{\max}$ is the maximum gain of each antenna element that has a typical value of 8 dBi, $\theta$ is the elevation angle between the dt-BS and the UAV-UE as shown in Fig. \ref{fig:sys},  $\theta_{\rm \rm dt}$ is the electrical downtilt angle and $\theta_{\rm 3dB}^{\rm dt}$ is the 3dB beamwidth. Moreover, $G_{\rm m}$ is the threshold for antenna nulls in antenna radiation pattern that typically has a value of $30$ dB. Note that $\theta_{\rm \rm dt} \in [-90^\circ, 0^\circ ]$ as measured from the horizontal axis.

The elevation angle $\theta$ with respect to the UAV-UE can be written as \cite{Cherif2025}
\begin{equation}
    \label{eq:elev}
    \theta = -\arcsin\left(\frac{h_{\rm uav}^{\rm dt}}{l}\right),
\end{equation}
where $h_{\rm uav}^{\rm dt}=h_{\rm uav}-h_{\rm dt}$ is the effective difference in altitudes between the antenna array of the dt-BS and the UAV-UE, and $l$ is the 3D distance separating the dt-BS from the UAV-UE.

The array factor of the downtilted ULA, $A_f^{\rm dt}$, is given by
\begin{equation}
    \label{eq:af}
    A_f^{\rm dt}(\theta)=\frac{1}{\sqrt{N_t}}\frac{\sin\left(\frac{N_t\pi}{2} ( \sin\theta-\sin\theta_{\rm dt})\right)}{\sin\left(\frac{\pi}{2} ( \sin\theta-\sin\theta_{\rm dt})\right)}.
\end{equation}
The array power gain in dB is denoted as $G_{f}^{\rm dt}(\theta)=10 \log_{10}(A_f^{\rm dt}(\theta)^2)$. Finally, the overall antenna array gain, in dB, at elevation angle $\theta$ from  dt-BS is given by 
\begin{equation}
    \label{eq:gain_tot}
    G_{\rm tot}^{\rm dt}(\theta)=G_e^{\rm dt}(\theta)+G_f^{\rm dt}(\theta).
\end{equation}

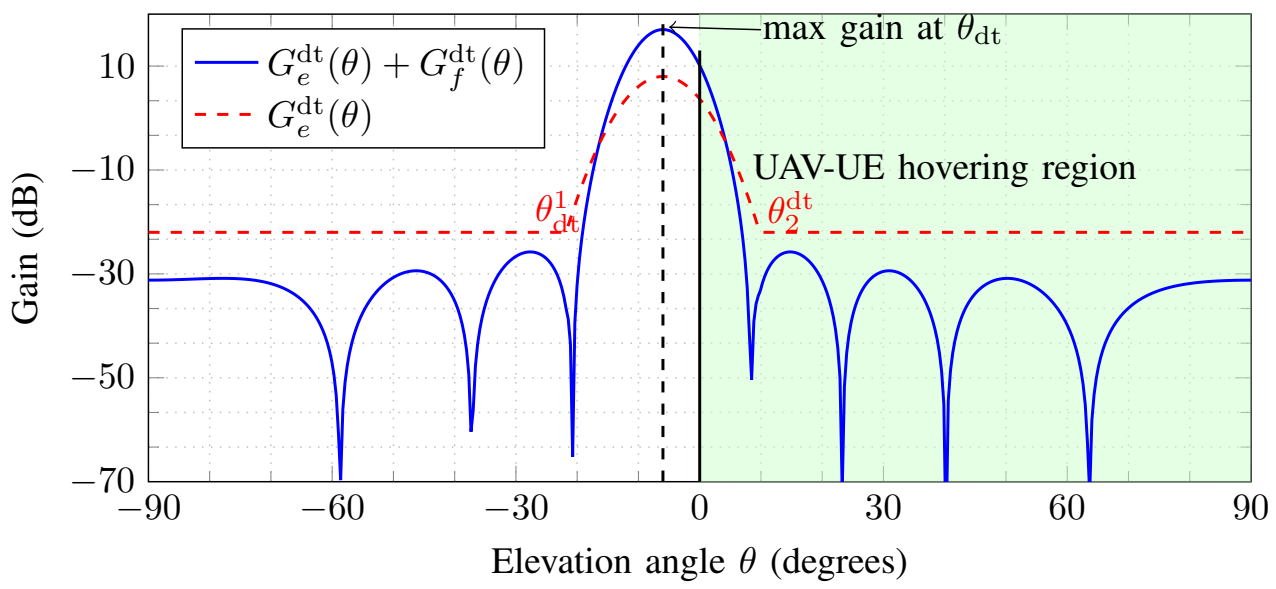
\begin{figure}
\centering
\begin{tikzpicture}[scale=0.7]
\begin{axis}[
    width=12cm,
    height=6cm,
    xlabel={Elevation angle $\theta$ (degrees)},
    ylabel={Gain (dB)},
    xmin=-90, xmax=90,
    ymin=-70, ymax=20,
    xtick={-90,-60,-30,0,30,60,90},
    ytick={-70,-50,-30,-10,10},
    grid=both,
    legend pos=north west,
    legend style={nodes={scale=1, transform shape}}, 
     minor tick num=2,
    xmajorgrids=true,
    xminorgrids=true,
    ymajorgrids=true,
    grid style=dotted,
    yminorgrids=true,
    legend cell align={left}
]

\fill[green!20, fill opacity=0.5] (axis cs:0,-70) rectangle (axis cs:90,20);

\pgfmathsetmacro{\thetaD}{-6} 
\pgfmathsetmacro{\N}{8}       
\pgfmathsetmacro{\HPBM}{10}   
\pgfmathsetmacro{\Gm}{30}     

\addplot [
    domain=-90:90,
    samples=500,
    thick, 
    color=blue
]
{10*ln(((1/sqrt(\N)) * sin(deg((sin(x) - sin(\thetaD))*\N*pi/2)) / (sin(deg((sin(x) - sin(\thetaD))*pi/2))))^2)/ln(10)
+ (8 - min(12*((x - \thetaD)/\HPBM)^2, \Gm))};
\addlegendentry{$G_e^{\rm dt}(\theta)+G_f^{\rm dt}(\theta)$}

\addplot [
    domain=-90:180,
    samples=500,
    thick, 
    dashed,
    color=red
]
{8 - min(12*((x - \thetaD)/\HPBM)^2, \Gm)};
\addlegendentry{$G_e^{\rm dt}(\theta)$}

\node (destination) at (axis cs:-23, -18){\color{red}$\theta_{\rm dt}^1$};
\node (destination) at (axis cs:15,-18){\color{red}$\theta_2^{\rm dt}$}; 
\addplot[thick, samples=50, dashed,domain=0:6,black] coordinates {(\thetaD,-70)(\thetaD,20)};  
\addplot[thick, samples=50, smooth,domain=0:6,black] coordinates {(0,-70)(0,13)};   
\node (destination) at (axis cs:30,17){max gain at $\theta_{\rm dt}$};
\draw [to-] (axis cs:-5, 17.5) --   (axis cs:10, 17);
\node (destination) at (axis cs:40,-10){UAV-UE hovering region};
\end{axis}
\end{tikzpicture}
\captionof{figure}{Antenna array gain of dt-BS for $(\theta_{\rm dt},\theta_{\rm 3dB}^{\rm dt})=(-6^{\circ},10^{\circ})$.}
\vspace{15pt}
\label{fig:elevation}
\end{figure}

In Fig. \ref{fig:elevation}, we plot the antenna element gain and the overall vertical antenna array gain in dB given by eq. (\ref{eq:antennavertical}) and eq. (\ref{eq:gain_tot}), respectively,  versus the elevation angle in [-90$^\circ$, 90$^\circ$]. It is worth mentioning that for a GUE, the elevation angle is always negative since the height of the GUE is smaller than that of the dt-BS. On the other hand, the elevation angle is always positive for the UAV-UE as the UAV-UE is expected to hover at altitudes higher than the BSs' height. Fig. \ref{fig:elevation} shows that the maximum radiation pattern gain can be attained when the user is at an elevation angle equal to the downtilt angle $\theta_{\rm dt}$. This means that the strength of the antenna radiation pattern in the direction where the GUE is located and the interference it causes where the UAV-UE is located will depend on the downtilt angle and the 3dB beamwidth set at the dt-BS. The antenna element gain, $G_e^{\rm dt}(\theta)$, attains its minimum value $G_{\max}-G_{\rm m}$ at two values of elevation angle, $\theta_1^{\rm dt}=-\sqrt{\frac{G_{\rm m}}{12}}\theta_{\rm 3dB}^{\rm dt}+\theta_{\rm dt}$ and $\theta_2^{\rm dt}=\sqrt{\frac{G_{\rm m}}{12}} \theta_{\rm 3dB}^{\rm dt}+\theta_{\rm dt}$  \cite{aquino2015}. Using this, we can express the overall antenna array gain in \eqref{eq:gain_tot} of dt-BS interference at the UAV-UE as shown in Table \ref{tab:tab1}. 

By plugging  (\ref{eq:elev}) into  $ G_{e}^{\rm dt}(\theta)$, the  antenna element gain as a function of the 3D distance $l$ between the UAV-UE and the interfering dt-BS attains its minimum value at $r_2^{\rm dt}=\frac{h_{\rm uav}^{\rm dt}}{\sin(\theta_2^{\rm dt})}$. Thus, we can express the overall interfering antenna pattern gain $G_{\rm tot}^{\rm dt}(l)$ in \eqref{eq:gain_tot} as a function of $l$, in linear scale, as shown in Table \ref{tab:tab2}, where
$G_{\rm ml}^{\rm dt}(l)=A_f^{\rm dt}(l)^2\times10^{\frac{G_{\max}}{10}-1.2\left(\frac{-\arcsin\left(\frac{h_{\rm uav}^{\rm dt}}{l}\right)-\theta_{\rm dt}}{\theta_{\rm 3dB}^{\rm dt}}\right)^2}$ and $G_{\rm sl}^{\rm dt}(l)=A_f^{\rm dt}(l)^2\times10^{\frac{G_{\max}-G_{\rm m}}{10}}$.

\begin{table}[t!]
\small
\centering
\renewcommand{\arraystretch}{1.5} 
\setlength{\tabcolsep}{8pt} 
\caption{Overall antenna array gain in dB of the dt-BS at the UAV-UE for different thresholds on $\theta_2^{\rm dt}$.}
\label{tab:tab1}

\begin{tabular}{c||l}
\hline \hline
\textbf{$\theta_2^{\rm dt}$} & \textbf{Overall antenna gain in dB: $G_{\rm tot}^{\rm dt}(\theta)$} \\ \hline \hline

$\theta_2^{\rm dt} \leqslant 0^{\circ}$ & 
$ G_f^{\rm dt}(\theta) + G_{\max} - G_{\rm m} $ \\ \hline

$\theta_2^{\rm dt} \geqslant 0^{\circ}$ & 
$
\begin{cases} 
G_f^{\rm dt}(\theta) + G_{\max} - 12\left(\frac{\theta-\theta_{\rm dt}}{\theta_{\rm 3dB}^{\rm dt}}\right)^2, & 0 \leqslant \theta < \theta_2^{\rm dt}, \\ 
G_f^{\rm dt}(\theta) + G_{\max} - G_{\rm m}, & \theta \geqslant \theta_2^{\rm dt}.
\end{cases}
$ \\ \hline \hline

\end{tabular}
\end{table}
\vspace{15pt}
\begin{table}[t!]
\small
\vspace{15pt}
\centering
\renewcommand{\arraystretch}{1.5} 
\setlength{\tabcolsep}{20pt} 
\caption{Overall antenna array gain in linear scale of the dt-BS at the UAV-UE as a function of $l$.}
\begin{tabular}{c||c||l}
\hline\hline
\textbf{$\theta_2^{\rm dt}$} & \textbf{$r_2^{\rm dt}$} & $G_{\rm tot}^{\rm dt}(l)$ \\ \hline\hline

$\theta_2^{\rm dt}\leqslant 0^{\circ}$ & NA & 
\parbox{1cm}{$\displaystyle G_{\rm sl}^{\rm dt}(l)$} \\ \hline

$\theta_2^{\rm dt}\geqslant 0^{\circ}$ & $\frac{h_{\rm uav}^{\rm dt}}{\sin(\theta_2^{\rm dt})}$ & 
\parbox{2cm}{$\displaystyle 
\begin{cases} 
G_{\rm sl}^{\rm dt}(l), & r < r^{\rm dt}_2, \\ 
G_{\rm ml}^{\rm dt}(l), & r \geqslant r^{\rm dt}_2.
\end{cases}
$} \\ \hline\hline
\end{tabular}
\label{tab:tab2}
\end{table}

\subsection{Antenna Gain Model of ut-BS}

Since we are collocating new uptilted antenna arrays on some dt-BSs for aerial coverage without changing the architecture of existing ground cellular service, i.e., downtilted antenna arrays, we enjoy more leverage in terms of which antennas we can deploy for UAV-UE coverage.  Since highly directional antennas are best suited to maximize the desired link gain between the serving ut-BS and the typical UAV-UE, we consider a UPA of $N_{\rm ut}$ directional antennas deployed at each ut-BS \cite{guo2021coverage}. The considered array results in an  antenna pattern with a half power beamwidth in the azimuth and elevation plane denoted by $\theta_{\rm ut}^{(a)}$ and $\theta_{\rm ut}^{(e)}$, respectively, and the mainlobe and sidelobe beam gains specified by $G_{\rm ut}$ and $g_{\rm ut}$, respectively (See \cite[Fig. 2]{venugopal2016device}). Similarly, we assume that the UAV-UE is equipped with a UPA  of $N_{\rm uav}$ antennas. The azimuth and elevation half-power beamwidth angles at the UAV-UE antenna array are denoted by $\theta_{\rm uav}^{(a)}$ and $\theta_{\rm uav}^{(e)}$, respectively, and the mainlobe and sidelobe beam gains are specified by $G_{\rm uav}$ and $g_{\rm uav}$, respectively. Table \ref{tab:upa} illustrates the parameters related to  UPAs deployed at the ut-BS and the UAV-UE with $\delta\in\{\rm ut, uav\}$ \cite{venugopal2016device}. 
\renewcommand{\arraystretch}{1.9} 
\begin{table}[t!]
\centering
\vspace{15pt}
\footnotesize
\caption{UPA radiation parameters at ut-BS and UAV-UE.}
\begin{tabular}{l l}
\hline \hline
Number of antenna elements                       & $N_\delta$                  \\ \hline \hline
 \parbox{4cm}{ Half power beamwidth, \\ $\theta^{(a)}_{\delta}=\theta^{(e)}_{\delta}$}
 & $\sqrt{\frac{3}{N_\delta}}$ \\ \hline
         Mainlobe gain, $G_\delta$        &     $N_\delta$                 \\ \hline
            Sidelobe gain, $g_{\delta}$     &     $  \frac{\sqrt{N_\delta}-\frac{\sqrt{3}}{2 \pi}N_\delta \sin\left( \frac{\sqrt{3}}{2 \sqrt{N_\delta}}\right)}{\sqrt{N_\delta}-\frac{\sqrt{3}}{2 \pi} \sin\left( \frac{\sqrt{3}}{2 \sqrt{N_\delta}}\right)} $                 \\ \hline \hline
\end{tabular}
\label{tab:upa}
\vspace{15pt}

\end{table}

The serving ut-BS and UAV-UE can align their mainlobes by leveraging the hierarchical beam search scheme \cite{guo2021coverage} to obtain the maximum gain of $G_{\rm uav} \times G_{\rm ut}$. \startmycolor Nevertheless, beam misalignment can occur as a result of UAV-UE mobility and/or extremely narrow mainlobe beamwidth of the ut-BS/UAV-UE \cite{kalamkar2022beam}, reducing the desired signal strength at the UAV-UE. A strong beam misalignment between the UAV-UE and the serving ut-BS is when the sidelobe of the selected beam of the ut-BS is aligned with its counterpart of the UAV-UE. More specifically, the probability of occurrence of beam misalignment depends on the speed of the UAV-UE  $v$, the number of antennas $N_{\rm ut}$ and $N_{\rm uav}$, and the duration between two synchronization signal bursts (SSBs) $t_{\rm SSB}$\footnote{For more details on beam management, readers are referred to \cite{kalamkar2022beam}}. Thus, the probability of beam misalignment on the $\delta$-th side  $p_{\rm bm}^{\rm \delta}$, $\delta\in\{\rm ut, uav\}$ is given by \cite[eqs. (3) and (4)]{kalamkar2022beam}
\begin{equation}
\label{eq:pbm}
  p_{\rm bm}^{\rm \delta} =1-\exp \left(-\frac{v \ t_{\rm SSB} \ {N_{\delta}}\ \sqrt{\lambda_{\rm ut}}}{\pi} \right).
\end{equation}
It follows that \textcolor{black}{the combined antenna gain of the serving ut-BS UAV-UE link}, i.e., $G_0$, is based on the beam misalignment at both the serving ut-BS and the typical UAV-UE and can be expressed by Table \ref{tab:ms}. \textcolor{black}{It is noteworthy that perfect time and frequency synchronization between the serving ut-BS and the UAV-UE is assumed, consistent with standard cellular system-level analysis, while the impact of residual synchronization impairments is left for future work.}
 
\begin{table}[t!]
\centering
\small
\caption{\textcolor{black}{Combined antenna gain of the serving ut-BS UAV-UE link} and their probabilities of occurrence.}
\begin{tabular}{l l l}
\hline \hline
 Gain $G_0$ & $\mathcal{P}_{G_0}$ \\ \hline \hline
      $G_{\rm uav} G_{\rm ut}$      & $(1-p_{\rm bm}^{\rm uav}) \times (1-p_{\rm bm}^{\rm ut})$              \\ \hline
        $G_{\rm uav} g_{\rm ut}$          &    $(1-p_{\rm bm}^{\rm uav})\times p_{\rm bm}^{\rm ut} $            \\ \hline
         $g_{\rm uav} G_{\rm ut}$       &       $p_{\rm bm}^{\rm uav}\times (1-p_{\rm bm}^{\rm ut})$        \\ \hline
      $g_{\rm uav} g_{\rm ut}$          &            $p_{\rm bm}^{\rm uav}\times p_{\rm bm}^{\rm ut}$     \\ \hline\hline
\end{tabular}
\label{tab:ms}
\vspace{15pt}
\end{table}

\stopmycolor

Further, we consider the non-serving ut-BSs and the interference they introduce at the UAV-UE as a result of their radiation patterns. We consider the azimuth and elevation angles of the propagation paths from non-serving ut-BSs at the UAV-UE to be uniformly distributed in  $[0,2\pi)$ and $[0,\pi]$ respectively \cite{guo2021coverage, venugopal2016device}\footnote{The interfering ut-BSs are assumed to be serving other UAV-UE. The probability that a typical  UAV-UE experiences an interfering ut-BS gain of $g_{\rm uav} G_{\rm ut}$ is $\left(1-\frac{\theta_{\rm uav}^{(a)}}{2\pi}\frac{\theta_{\rm uav}^{(e)}}{\pi}\right)\frac{\theta_{\rm ut}^{(a)}}{2\pi}\frac{\theta_{\rm ut}^{(e)}}{\pi}$  according to the uniform distribution of the angles (since the UAV-UE is experiencing the mainlobe gain of the interfering ut-BS through its own sidelobe region). Other probabilities of interference tabulated in Table VI can be similarly computed.}. As a result, the total antenna array gain at the UAV-UE from an interfering ut-BS is a discrete random variable $\Upsilon_{\Omega}\in\Omega=\{G_{\rm uav} G_{\rm ut},\ G_{\rm uav} g_{\rm ut},\ g_{\rm uav} G_{\rm ut},\ g_{\rm uav} g_{\rm ut}\}$ with probability of occurrence $\mathcal{P}_{\Upsilon_{\Omega}}$  summarized in Table \ref{tab:probint}.
\begin{table}[t!]
\centering
\small
\caption{Interfering antenna gain and corresponding probability of occurrence for an interfering link between a non-serving ut-BS and UAV-UE.}
\begin{tabular}{l l l}
\hline \hline
 Gain $\Upsilon_{\Omega}$ & $\mathcal{P}_{\Upsilon_{\Omega}}$ \\ \hline \hline
      $G_{\rm uav} G_{\rm ut}$      & $\left(\frac{\theta_{\rm uav}^{(a)}}{2\pi}\frac{\theta_{\rm uav}^{(e)}}{\pi}\right)\times\left(\frac{\theta_{\rm ut}^{(a)}}{2\pi}\frac{\theta_{\rm ut}^{(e)}}{\pi}\right) $               \\ \hline
        $G_{\rm uav} g_{\rm ut}$          & $\left(\frac{\theta_{\rm uav}^{(a)}}{2\pi}\frac{\theta_{\rm uav}^{(e)}}{\pi}\right)\times\left(1-\frac{\theta_{\rm ut}^{(a)}}{2\pi}\frac{\theta_{\rm ut}^{(e)}}{\pi}\right) $                \\ \hline
         $g_{\rm uav} G_{\rm ut}$       &   $\left(1-\frac{\theta_{\rm uav}^{(a)}}{2\pi}\frac{\theta_{\rm uav}^{(e)}}{\pi}\right)\times\left(\frac{\theta_{\rm ut}^{(a)}}{2\pi}\frac{\theta_{\rm ut}^{(e)}}{\pi}\right) $              \\ \hline
      $g_{\rm uav} g_{\rm ut}$          &    $\left(1-\frac{\theta_{\rm uav}^{(a)}}{2\pi}\frac{\theta_{\rm uav}^{(e)}}{\pi}\right)\times\left(1-\frac{\theta_{\rm ut}^{(a)}}{2\pi}\frac{\theta_{\rm ut}^{(e)}}{\pi}\right) $             \\ \hline\hline
\end{tabular}
\label{tab:probint}
\vspace{15pt}
\end{table}

\subsection{Channel Model}
In this subsection, we focus on the channels between the UAV-UE and the ut-BSs and dt-BSs. \textcolor{black}{As mentioned above, the UAV-UE is assumed to connect only to ut-BS.} The channels connecting the UAV-UE and the dt-BSs, as well as the UAV-UE and non-serving ut-BSs are considered to be the interfering channels at the UAV-UE. As shown in Fig. \ref{fig:sys}, depending on the downtilt angle at the dt-BSs, the UAV-UE might experience interference from the sidelobe or from the reflection of the mainlobe of the radiation pattern of the dt-BSs. 

\subsubsection{Channel Model for ut-BS UAV-UE Link}
We consider that the link between ut-BS and the UAV-UE can be LoS (L) or NLoS (N). The LoS and NLoS links have different path loss exponents, denoted by $\alpha_{\rm L}$ and $\alpha_{\rm N}$ respectively, and experience penetration losses of $\mu_{\rm L}$ and $\mu_{\rm N}$, respectively \cite{alhouranilet}. \startmycolor We assume that the channel modeling each link (whether LoS or NLoS) undergoes Nakagami-m fading with shape parameter $m$. Thus, the channel gain for  $\tau$-th link, $\tau \in \{\rm L,N\}$, between the UAV-UE and the $i$-th ut-BS follows Gamma distribution, i.e., $q^{\tau}_{{\rm ut},i}\sim \text{Gamma}\left(m_{\tau},\frac{1}{m_{\tau}}\right)$\footnote{\startmycolor Nakagami-m fading is a versatile fading model that encompasses other models such as Rician-K fading by setting  $m=(K+1)^2/2K+1$ and Rayleigh fading by setting  $m=1$.\stopmycolor}\stopmycolor. 

The channel between the ut-BS and the UAV-UE is LoS with probability  $\mathcal{P}^{\rm L}$ expressed as \cite[eq. (2)]{kim2019multi}
\begin{equation}
    \label{eq:plos1}
    \mathcal{P}^{\rm L}(r)\!=\!\left(\!1\!-\!\frac{\sqrt{2 \pi}\xi}{h_{\rm uav}^{\rm ut}}\left|\!\mathcal{Q}\!\left(\!\frac{h_{\rm uav}}{\xi}\!\right)\!\!-\!\!\mathcal{Q}\!\left(\!\frac{h_{\rm ut}}{\xi}\!\right)\!\right|\!\right)^{\!\!\!\sqrt{\frac{ \nu \kappa}{1000^2}}r},
\end{equation}
where $\xi$, $\nu$, and $\kappa$ are environment-related parameters \cite{alhouranilet}. More specifically, $\nu$ is the ratio of the area covered by buildings to the total area, $\kappa$ is the mean number of buildings in km$^2$,  $\xi$ is the average height of buildings, $\mathcal{Q} (\cdot)$ is the Q-function, $r$ is the 2D distance between the ut-BS and the UAV-UE, and $h_{\rm uav}^{\rm ut}=h_{\rm uav}-h_{\rm ut}$ is the effective height difference between the UAV-UE and the ut-BS. We can rewrite the probability of LoS  using the exponential function definition $a^{b x}=\exp\left({\log(a) b x}\right)$ as
\begin{equation}
    \label{eq:plos}
    \mathcal{P}^{\rm L}(r)=\exp\left(\mathcal{K} \times r\right),
\end{equation}
where $\mathcal{K}=\log\left(1-\frac{\sqrt{2 \pi}\xi}{h_{\rm uav}^{\rm ut}}\left|\mathcal{Q}\left(\frac{h_{\rm uav}}{\xi}\right)-\mathcal{Q}\left(\frac{h_{\rm ut}}{\xi}\right)\right|\right) \sqrt{\frac{ \nu \kappa}{1000^2}}$.

The ut-BS transmission to the UAV-UE can be either LoS or NLoS, depending on their location. Thus, invoking the thinning theorem \cite{haenggi2012stochastic}, the set
of ut-BS can be split into two independent
inhomogeneous PPPs, where the LoS ut-BS tier $\Phi_{\rm UT}^{\rm L}$ has density $\lambda_{\rm ut}\times \mathcal{P}^{\rm L}(r)$ and the NLoS ut-BS tier $\Phi_{\rm UT}^{\rm N}$ has density $\lambda_{\rm ut}\times \mathcal{P}^{\rm N}(r)$. Please note that $\mathcal{P}^{\rm N}(r)=1-\mathcal{P}^{\rm L}(r)$.

The received power at the UAV-UE through the  $\tau$-th link, where $\tau \in \{\rm L,N\}$,  with respect to the $i$-th ut-BS, \textcolor{black}{separated by the 3D distance  $r_{i, \tau}$}, is given by
\startmycolor
\begin{equation}
    \label{eq:prlosnlos}
    \textcolor{black}{P^{\rm ut,\tau}_i=P_{\rm ut}\left[\frac{\lambda}{4 \pi}\right]^2 \mu_{\tau}\ \mathcal{G}\ q^{\tau}_{{\rm ut},i} r_{i, \tau}^{-\alpha_{\tau}},}
    \end{equation}
\stopmycolor
where $\lambda$ is the wavelength of the carrier frequency. \textcolor{black}{ It is worth highlighting that $P^{\rm ut,\tau}_i$ is the desired  signal power when received from the serving ut-BS with $\mathcal{G} \ =G_0$ (defined in Table \ref{tab:ms}), and the interference signal power when received from non-serving ut-BSs with $\mathcal{G} \ =\Upsilon_{\Omega}$ (defined in  Table \ref{tab:probint}).}

\subsubsection{GR-based Channel Model for dt-BS UAV-UE Link}

Most existing works in the literature assume that interference at the UAV-UE from a dt-BS arises primarily through the sidelobes, while largely overlooking the significant contribution from the mainlobe. In reality, the mainlobe of a dt-BS, when reflected off the ground, can propagate toward the UAV-UE, introducing substantial interference that can exceed that caused by sidelobe interference. This phenomenon is particularly critical in scenarios involving low-altitude UAV operation or environments with reflective ground surfaces, such as urban areas or water bodies, where the reflection of signals from the ground is more pronounced. To address this limitation, we  account for both the direct sidelobe interference  and the interference caused by ground-reflected mainlobe of the radiation pattern of the dt-BS, enabling a more accurate characterization of the dt-BS UAV-UE channel.

\begin{figure}[t!]
	\centering
	\includegraphics[ width=0.4\textwidth, height=5cm,]{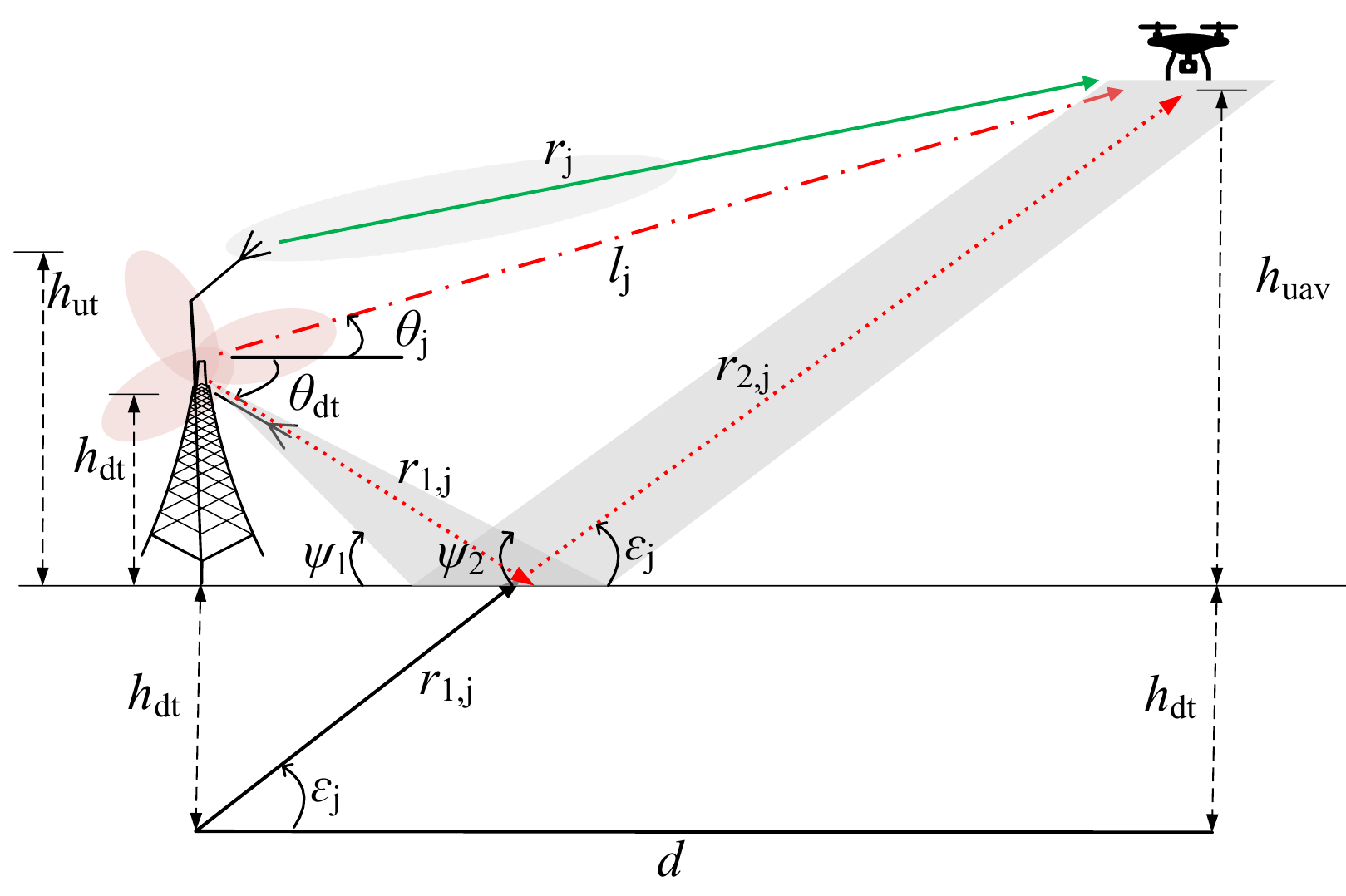}
	\caption{ System geometry. }
	\label{fig:geometry_sys}
\vspace{7pt}
\end{figure}

\startmycolor
More specifically, the channel between the dt-BS and the UAV-UE accounts for both distance-dependent path loss as well as sidelobe and mainlobe components of radiation pattern of the dt-BS. Following the models in \cite{goddemeier2012role,chowdhury2021ensuring}, we adopt a height-dependent variant of the ground-reflection radiation pattern model \cite[eq.~(6)]{chowdhury2021ensuring}. Let $l_j$ denote the direct 3D distance between the $j$-th dt-BS and the UAV-UE, while $r_{1,j}$ and $r_{2,j}$ represent the respective lengths of the incident and reflected beams via the ground (see Fig.~\ref{fig:geometry_sys}). The interference power received at the UAV-UE located at height $h_{\rm uav}$  from the  $j$-th dt-BS is then given by
\begin{eqnarray}
      \label{eq:pr_gr}
        \!\!\!\!\!\!\!\!\!\!\!\!\!\!\!&&P_{j}^{\rm dt} = P_{\rm dt} \left[\frac{\lambda}{4 \pi}\right]^2q_{{\rm dt},j}\Psi_{\omega}\\ \nonumber
     \!\!\!\!\!&&\times\left | \frac{\sqrt{G_{\rm tot}^{\rm dt}(l_j)}}{l_j}+ \mathds{1}_{\psi_1}^{\psi_2}(\epsilon_j)\frac{R(\epsilon_j)\sqrt{G_{\rm ref}^{\rm dt}(h_{\rm uav},\epsilon_j)} e^{i \Delta \phi_j}}{r_{1,j}+r_{2,j}}\right |^{\alpha(h_{\rm uav})},
\end{eqnarray}
where the channel gain between the $j$-th dt-BS and the UAV-UE,  $q_{{\rm dt}, j}$, is exponentially-distributed as $q_{{\rm dt}, j}\sim\text{exp}\left(1\right)$, with its probability density function  (PDF) expressed as $f_{{q_{{\rm dt}, j}}}(x)=e^{-x}$.  Further
$\Psi_{\omega} \in W=\{G_{\rm uav},g_{\rm uav}\}$ is the antenna gain of the UAV-UE in the direction of the $j$-th interfering dt-BS with a probability of occurrence $\mathcal{P}_{\Psi_{\omega}}\in \Big\{\left(\frac{\theta_{\rm uav}^{(a)}}{2\pi}\frac{\theta_{\rm uav}^{(e)}}{\pi}\right),1-\left(\frac{\theta_{\rm uav}^{(a)}}{2\pi}\frac{\theta_{\rm uav}^{(e)}}{\pi}\right)\Big\}$, respectively, and \textcolor{black}{$G_{\rm tot}^{\rm dt}(l_j)$ is the dt-BS sidelobe antenna gain (refer to Table \ref{tab:tab2})}. Moreover, $R(\epsilon_j)$ is the GR coefficient at the angle of reflection $\epsilon_j$ with respect to the ground plane, and \textcolor{black}{$G_{\rm ref}^{\rm dt}$ is the height-dependent antenna gain for the reflected beam from the dt-BS.} Moreover, $\Delta \phi_j=(r_{1,j}+r_{2,j})-l_j$ is the phase difference of the reflected and direct signals. \textcolor{black}{We define $\mathds{1}_{\psi_1}^{\psi_2}$ as the indicator function with $\psi_1 = \theta_{\rm dt}-0.5\times\theta_{\rm 3dB}^{\rm dt}$ and $\psi_2 = \theta_{\rm dt}+0.5\times\theta_{\rm 3dB}^{\rm dt}$ (See Fig. \ref{fig:geometry_sys}).} It is worth highlighting that the GR from the mainlobe of the  dt-BS will have an effective impact between $\psi_1$ and $\psi_2$ \cite[Theorem 1]{chowdhury2021ensuring} as shown in Fig. \ref{fig:geometry_sys}.

 It should be noted that, in realistic wireless channels, the received signal contains multiple reflected and scattered components in addition to the LoS path. This small-scale fading is explicitly modeled in the considered channel model in \eqref{eq:pr_gr} through the variable $q_{\rm dt, j}$. The GR model in \cite[eq.(6)]{chowdhury20203d} and \cite[eq. (10)]{goddemeier2012role} only modifies the antenna radiation pattern to include not just the direct radiation pattern of the dt-BS experienced by the UAV-UE (i.e. $G_{\rm tot}^{\rm dt}(l_j)$), but also the ground reflected radiation pattern capturing the reflection of the mainlobe of the dt-BS radiation pattern through $G_{\rm ref}^{\rm dt}(h_{\rm uav}, \epsilon_j)$ experienced by the UAV-UE. The multipath components are still captured by the small-scale fading coefficient ${q_{{\rm dt}, j}}$, which statistically aggregates the effect of multiple scatterers. This is standard in wireless channel models that express the overall channel as a product of the antenna radiation pattern and small-scale fading. Note that when $R(\epsilon_j)=0$, the model in \eqref{eq:pr_gr} simplifies to the scenario where the ground reflection of the mainlobe of the dt-BS radiation pattern is not considered to contribute to the overall dt-BS UAV-UE channel, confirming that the model in \eqref{eq:pr_gr} is just an enhancement of the existing spatial channel models. \stopmycolor

 The path loss exponent for the link separating the interfering dt-BS and the UAV-UE is assumed to be height dependent \cite[eq. (11)]{goddemeier2012role} and it is specified as \stopmycolor
\begin{equation}
    \label{eq:alphah}
    \alpha({h_{\rm uav}})=
    \begin{cases}
        \alpha_0-\frac{h_{\rm uav}}{h_{\rm dt}}(\alpha_0-2),&h_{\rm uav} < 2 h_{\rm dt}\\
        2,& h_{\rm uav}  \geqslant2 h_{\rm dt}
    \end{cases}
\end{equation}
where $\alpha_0=3.5$ is the maximum possible attenuation coefficient \cite{goddemeier2012role}. Please note that with a typical height of a dt-BS that is around 25 meters, and a minimum allowed UAV-UE altitude (for safety regulations) of 50 meters \cite{3gpp777}, the height-dependent path loss exponent in \eqref{eq:alphah} always equals the free space path loss exponent, which is $2$.


Note that  the GR coefficient of a cross-polarized antenna is derived as $R(\epsilon_j)=\frac{R_{\rm H}(\epsilon_j)-R_{\rm V}(\epsilon_j)}{2}$ which depends on the relative ground permittivity  (e.g, for medium dry ground: $\epsilon_p = 15$), and the reflection coefficients for horizontal and vertical linear polarizations denoted as $R_{\rm H}(\epsilon_j)$ and $R_{\rm V}(\epsilon_j)$, respectively \cite{najibi2013physical}, \cite[eqs.(90, 91)]{jordan1951electromagnetic}, \cite{tr138901}.

The height-dependent antenna gain of the reflected mainlobe $G_{\rm ref}^{\rm dt}$ depends on the angle of reflection $\epsilon_j$ with respect to the ground plane, and it is specified as
\begin{equation}
    \label{eq:gain_ref}
    G_{\rm ref}^{\rm dt}(h_{\rm uav},\epsilon_j)=    \begin{cases}
        G_{\rm tot}^{\rm dt}(\epsilon_j), &h_{\rm uav}<h_t\\
        \frac{G_{\rm tot}^{\rm dt}(\epsilon_j)}{2}, & h_t \leqslant h_{\rm uav}\leqslant 2 h_t\\
        \frac{G_{\rm tot}^{\rm dt}(\epsilon_j)}{2} -\frac{h_{\rm uav}}{2 h_{\max}}\\
       \times(G_{\rm tot}^{\rm dt}(\epsilon_j)-1), & 2 h_t \leqslant h_{\rm uav}\leqslant h_{\max}\\
        \frac{1}{2}, & h_{\rm uav}\geqslant h_{\max},
    \end{cases}
\end{equation}
where $ G_{\rm tot}^{\rm dt}(\epsilon_j)$ is defined in Table \ref{tab:tab1}, $h_{\max}=500$ m \cite{goddemeier2012role} and $h_t=2 \times h_{\rm dt}+2$.

It is essential to derive the total distance traveled by the reflected path, i.e., $r_{1,j}+r_{2,j}$, along with the angle of reflection $\epsilon_j$ as a function of the network parameters, mainly dt-BS UAV-UE distance $l_j$. We illustrate the GR model involving the reflection of the  mainlobe   from dt-BS in Fig. \ref{fig:geometry_sys}. Using this, we formulate the reflection-dependent parameters as 
\begin{equation}
    \label{eq:r1r2}
    r_{1,j}+r_{2,j} = \sqrt{l_j^2+4 h_{\rm uav} h_{\rm dt}}
\end{equation}
\begin{equation}
    \label{eq:refangle}
   \epsilon_j = \arcsin \left(\frac{h_{\rm uav} +h_{\rm dt}}{\sqrt{l_j^2+4 h_{\rm uav} h_{\rm dt}}}\right)
\end{equation}
\textit{Proof:} The proof is postponed to Appendix \ref{ap:geometry}.

Please note that eq. \eqref{eq:pr_gr} can now be expressed solely as a function of the distance $l_j$ separating the $j$-th dt-BS and the typical UAV-UE by substituting eqs. \eqref{eq:gain_ref}, \eqref{eq:r1r2}, and \eqref{eq:refangle}.

\section{Coverage Probability Analysis}
In this section, we derive the coverage probability expressions for a typical UAV-UE operating in the considered overlaid terrestrial network with ut-BSs and dt-BSs, taking into account the interference from non-serving ut-BSs and the interference from the dt-BSs due to sidelobes and ground-reflected mainlobe. Assuming an interference-limited scenario, the impact of noise is neglected for simplicity, though it can be seamlessly incorporated into the calculations if needed.
\subsection{UAV-UE Association Rule}
\startmycolor
In this paper, we assume that the UAV-UE is connected to the ut-BS that provides the strongest long-term averaged received power \cite{andrews}. Assuming that the average power
of small-scale fading in all channels is one, i.e., $\mathbb{E}[l_{\rm ut,i}^{\rm L}]=\mathbb{E}[l_{\rm ut,i}^{\rm N}]=1$, the serving ut-BS would be the nearest ut-BS from tier $\Phi_{\rm UT}^{\rm L}$ or $\Phi_{\rm UT}^{\rm N}$. Mathematically, the serving ut-BS index denoted by $x^*$ is given as
\stopmycolor
\begin{equation}
    \label{eq:x*}
    x^*=\underset{x_i}{\text{arg max}} \{ \mu_{\rm L} d_{\rm L}^{-\alpha_{\rm L}},\mu_{\rm N} d_{\rm N}^{-\alpha_{\rm N}}\}
\end{equation}
where $d_{\rm L}=\underset{x_i \in \Phi_{\rm UT}^{\rm L}}{\min} r_{i,\rm L}$ and $d_{\rm N}=\underset{x_i \in \Phi_{\rm UT}^{\rm N}}{\min} r_{i,\rm N}$.


Considering that the UAV-UE operates in an interference-limited environment,  the downlink SIR  $\gamma$ at the UAV-UE, assuming that the UAV-UE and the serving ut-BS align their mainlobes to maximize the gain, is given by
\begin{equation}
\label{eq:SIR}
    \gamma = 
    \begin{cases}
       \frac{P_{\rm ut} \left[\frac{\lambda}{4 \pi}\right]^2\mu_{\rm L} G_{0}  l^{\rm L}_{{\rm ut},x^*} d_{\rm L}^{-\alpha_{\rm L}}}{I}, & x^* \in \Phi_{\rm UT}^{\rm L},\\
        \frac{P_{\rm ut} \left[\frac{\lambda}{4 \pi}\right]^2\mu_{\rm N} G_{0}  l^{\rm N}_{{\rm ut},x^*} d_{\rm N}^{-\alpha_{\rm N}}}{I}, & x^* \in \Phi_{\rm UT}^{\rm N},
    \end{cases}
\end{equation}
where $I$ denotes the total interference
power from non-serving LoS ut-BSs, non-serving NLoS ut-BSs, and the dt-BSs. It is specified as
$ I=I_{\rm UT}^{\rm L}+I_{\rm UT}^{\rm N}+I_{\rm DT}$, 
where $I_{\rm UT}^{\rm \tau}$ with $\tau \in \{\rm L, N\}$ is given by
{
\begin{align}
    \label{eq:INL}
    I_{\rm UT}^{\tau}&= \underset{x_i \in \Phi_{\rm UT}^{\tau} \backslash x^*}{\sum}P^{\rm ut,\tau}_i,\nonumber\\
    &=\underset{x_i \in \Phi_{\rm UT}^{\tau} \backslash x^*}{\sum} P_{\rm ut} \left[\frac{\lambda}{4 \pi}\right]^2\mu_{\tau} \Upsilon_{\Omega} q^{\tau}_{{\rm ut},i} r_{i, \tau}^{-\alpha_{\tau}},
\end{align}}
where $\Upsilon_{\Omega}$ is the gain of the non-serving interfering link with a probability of occurrence given in Table \ref{tab:probint}. Moreover, the dt-BS interference is formulated using \eqref{eq:pr_gr}, \eqref{eq:r1r2} and \eqref{eq:refangle} as 
{
\begin{eqnarray}
      \label{eq:IGR}
        \!\!\!\!\!\!\!\!\!\!\!\!\!\!\!&&I_{\rm DT} =\underset{x_j \in \Phi_{\rm DT}}{\sum} P_{j}^{\rm dt}= \underset{x_j \in \Phi_{\rm DT}}{\sum}P_{\rm dt} \left[\frac{\lambda}{4 \pi}\right]^2q_{{\rm dt},j} \Psi_{\omega}\\ \nonumber
     \!\!\!\!\!&&\times\left | \frac{G_{\rm tot}^{\rm dt}(l_j)}{l_j}+ \mathds{1}_{\psi_1}^{\psi_2}(l_j)\frac{R(l_j)G_{\rm ref}^{\rm dt}(l_j) e^{i \Delta \phi_j}}{r_{1,j}+r_{2,j}}\right |^{\alpha(h_{\rm uav})},
\end{eqnarray}}

\subsection{Distance Distributions}
In this section, we derive the relevant distance distributions that are crucial for the computation of the total coverage probability of the UAV-UE as well as the Laplace transform of the aggregate interference.
\begin{lemma}
The PDF and the cumulative distribution function (CDF) of the distance of the UAV-UE from the nearest LoS ut-BS, i.e. $d_{\rm L}$ defined after \eqref{eq:x*}, are given as 
\begin{eqnarray}
\label{eq:frl}
    &&f_{d_{\rm L}}(r)=2 \pi \lambda_{\rm ut} r \exp\left(\mathcal{K}\sqrt{r^2-(h_{\rm uav}^{\rm ut})^2}\right) \exp\Bigg(-2 \pi \lambda_{\rm ut} \nonumber \\
 && \times \frac{\exp\left(\mathcal{K}\sqrt{r^2-(h_{\rm uav}^{\rm ut})^2}\right)\left(\mathcal{K}\sqrt{r^2-(h_{\rm uav}^{\rm ut})^2}-1\right) + 1}{\mathcal{K}^2}\Bigg).\nonumber\\
\end{eqnarray}  
\begin{eqnarray}
\label{eq:Frl}
    &&F_{d_{\rm L}}(r)=1-\exp\Bigg(-2 \pi \lambda_{\rm ut}\exp\left(\mathcal{K}\sqrt{r^2-(h_{\rm uav}^{\rm ut})^2}\right)\nonumber \\
 && \times \frac{\left(\mathcal{K}\sqrt{r^2-(h_{\rm uav}^{\rm ut})^2}-1\right) }{\mathcal{K}^2}+\frac{2 \pi \lambda_{\rm ut}}{\mathcal{K}^2}\Bigg).
\end{eqnarray}
\end{lemma}

\textit{Proof:} The proof is postponed to Appendix \ref{ap:frl}.
\begin{lemma}
Similarly, the PDF and the CDF of the distance of the UAV-UE from the nearest NLoS ut-BS, $d_{\rm N}$, are given as
\begin{align}
\label{eq:frn}
    &f_{d_{\rm N}}(r)=2 \pi \lambda_{\rm ut} r \left(1-\exp\left(\mathcal{K}\sqrt{r^2-(h_{\rm uav}^{\rm ut})^2}\right)\right)  \nonumber \\
 & \!\!\!\times \exp\Bigg(-2 \pi \lambda_{\rm ut}\Bigg(\!\! \frac{r^2-(h_{\rm uav}^{\rm ut})^2}{2}-\exp\left(\!\mathcal{K}\sqrt{r^2-(h_{\rm uav}^{\rm ut})^2}\right)\nonumber\\
 &\!\!\!\times\frac{\!\!\left(\!\mathcal{K}\sqrt{r^2-(h_{\rm uav}^{\rm ut})^2}-1\right)}{\mathcal{K}^2}-\frac{1}{\mathcal{K}^2}\!\Bigg)\!\!\Bigg).
\end{align}
\begin{align}
\label{eq:Frn}
    &F_{d_{\rm N}}(r)=1-\exp\Bigg(-2 \pi \lambda_{\rm ut} \times \Bigg(\!\! \frac{r^2-(h_{\rm uav}^{\rm ut})^2}{2}  \\
 & \ \!\!\!-\frac{\exp\left(\mathcal{K}\sqrt{r^2-(h_{\rm uav}^{\rm ut})^2}\right)\left(\mathcal{K}\sqrt{r^2-(h_{\rm uav}^{\rm ut})^2}-1\right) + 1}{\mathcal{K}^2}\!\Bigg)\!\Bigg).\nonumber
\end{align}  
\end{lemma}
\textit{Proof:} Applying similar steps as in Appendix \ref{ap:frl} yields the result above.

\textcolor{black}{Moreover, the PDF of the distance separating the UAV-UE and the $j$-th dt-BS, i.e., $l_j$, is given by }\cite{nesrineglobecom}
\begin{equation}
    \label{eq:fdt}
    f_{l_j}(r)= 2 \pi \lambda_{\rm dt} r \exp \left(-2 \pi \lambda_{\rm dt}\sqrt{r^2-(h_{\rm uav}^{\rm dt})^2} \right), r\geqslant h_{\rm uav}^{\rm dt}.
\end{equation}
The derivation of the PDFs of the distances, i.e., $d_{\rm L}$ and $d_{\rm N}$, is crucial for the calculation of the likelihood that the serving ut-BS belongs to the LoS or NLoS ut-BS tier. We will now derive exact expressions of the distances of the UAV-UE to the closest interfering LoS ut-BS and NLoS ut-BS denoted by $d_{\rm L|N}(r)$ and $d_{\rm N|L}(r)$, assuming that the serving ut-BS belongs to NLoS or LoS ut-BS tiers, respectively, in the next subsection. Characterizing the distance to the closest interfering ut-BS will allow us to utilize it as a lower integration limit for the distance of all other interfering ut-BSs in the interference analysis.

\subsubsection{Distance to the Closest Interfering LoS/NLoS ut-BS}

Depending on the association of the typical UAV-UE to a LoS or NLoS serving ut-BS, the distance interval over which an interferer (LoS or NLoS ut-BS) may exist differs. Assuming the UAV-UE is connected to a NLoS ut-BS, the distance from the closest interfering LoS ut-BS to the UAV-UE is specified as
\begin{equation}
\label{eq:dln}
    d_{\rm L|N}(r)=\left(\frac{\mu_{\rm L}}{\mu_{\rm N}}\right)^{\frac{1}{\alpha_{\rm L}}} r^{\frac{\alpha_{\rm N}}{\alpha_{\rm L}}}.
\end{equation}
\textit{Proof:} Assuming the serving ut-BS is in NLoS with the typical UAV-UE with distance $r$, the average received power at the UAV-UE is given as $P_{\rm ut}\left[\frac{\lambda}{4 \pi}\right]^2 \mu_{\rm N} G_0  r^{-\alpha_{\rm N}}$. The distance with respect to the nearest interfering LoS ut-BS is denoted by $d_{\rm L|N}(r)$. Thus, by solving $P_{\rm ut} \left[\frac{\lambda}{4 \pi}\right]^2\mu_{\rm N} G_0  r^{-\alpha_{\rm N}}\geqslant P_{\rm ut}\left[\frac{\lambda}{4 \pi}\right]^2 \mu_{\rm L} G_0  d_{\rm L|N}(r)^{-\alpha_{\rm L}}$, we obtain eq. \eqref{eq:dln}.

On the other hand, the distance of the closest interfering NLoS ut-BS to the typical UAV-UE, assuming that the UAV-UE is connected to a LoS ut-BS, is given by
\begin{equation}
\label{eq:dnl}
    d_{\rm N|L}(r)=
    \begin{cases}
        h_{\rm uav}^{\rm ut},& h_{\rm uav}^{\rm ut}\leqslant r< \left(\frac{\mu_{\rm L}}{\mu_{\rm N}}\right)^{\frac{1}{\alpha_{\rm L}}} (h_{\rm uav}^{\rm ut})^{\frac{\alpha_{\rm N}}{\alpha_{\rm L}}}\\
        \left(\frac{\mu_{\rm N}}{\mu_{\rm L}}\right)^{\frac{1}{\alpha_{\rm N}}} r^{\frac{\alpha_{\rm L}}{\alpha_{\rm N}}}, &r\geqslant \left(\frac{\mu_{\rm L}}{\mu_{\rm N}}\right)^{\frac{1}{\alpha_{\rm L}}} (h_{\rm uav}^{\rm ut})^{\frac{\alpha_{\rm N}}{\alpha_{\rm L}}}. 
    \end{cases}
\end{equation}

\textit{Proof:} It follows the same steps as the proof of \eqref{eq:dln}.

It is important to note that when the interfering ut-BS belongs to the same tier as the serving ut-BS, i.e., the interfering ut-BS is in LoS tier given the serving ut-BS is in LoS tier or the interfering ut-BS is in NLoS tier given the serving ut-BS is also in NLoS tier, then the distance from the closest interfering ut-BS from that tier is always at least  $r$ \cite{andrews}. More specifically, $d_{\rm L|L}(r)=r$ and $d_{\rm N|N}(r)=r$. This remark will be used subsequently in the derivation of the coverage probability of the typical UAV-UE.
\subsubsection{Association Probabilities with LoS and NLoS ut-BS}
The typical UAV-UE is connected to either LoS or NLoS ut-BS with probabilities specified in the following propositions.
\begin{proposition}
The probability that the typical UAV-UE connects to a NLoS ut-BS is given by 
\begin{eqnarray}
     \label{eq:al}
  {\rm A^N}&=&\mathbb{P}\left[\mu_{\rm L}d_{\rm L}^{-\alpha_{\rm L}}\leqslant \mu_{\rm N}d_{\rm N}^{-\alpha_{\rm N}}\right]=\mathbb{P}\left[d_{\rm L}\geqslant d_{\rm L|N}(r)\right],\nonumber\\
  &=&\int_{h_{\rm uav}^{\rm ut}}^{\infty}(1-F_{d_{\rm L}}(d_{\rm L|N}(r))) f_{d_{\rm N}}(r)dr,
\end{eqnarray}
 where the last equality follows from averaging the complementary CDF over $d_{\rm N}$.
\end{proposition}
\begin{proposition}
By invoking the law of total probability, the likelihood that the typical UAV-UE associates with LoS ut-BS is specified by $ {\rm A^L}=1- {\rm A^N}$.
\end{proposition}

\subsubsection{Distance Distribution with respect to the Serving LoS and NLoS ut-BS}

The distribution of the distance between the typical UAV-UE and the serving LoS ut-BS defined by $\Tilde{d}_{\rm L}$ is specified as
\begin{equation}
    \label{eq:dlhat}
    f_{\Tilde{d}_{\rm L}}(r)=\frac{1-F_{d_{\rm N}}(d_{\rm N|L}(r))}{{\rm A^L}}\times f_{d_{\rm L}}(r).
\end{equation}
Similarly, the PDF of the distance between the typical UAV-UE and the serving NLoS ut-BS, $\Tilde{d}_{\rm N}$, is given by
\begin{equation}
    \label{eq:dnhat}
    f_{\Tilde{d}_{\rm N}}(r)=\frac{1-F_{d_{\rm L}}(d_{\rm L|N}(r))}{{\rm A^N}}\times f_{d_{\rm N}}(r).
\end{equation}
\textit{Proof:} We start by invoking Bayes's rule on  the CDF of $\Tilde{d}_{\rm L}$  as
\begin{align}
    \label{eq:dnhatpr}
    F_{\Tilde{d}_{\rm L}}(r)=&\frac{\mathbb{P}\left[d_{\rm L}<r;\text{serving ut-BS is LoS}\right]}{{\rm A^L}},\nonumber\\
    =&\frac{\mathbb{P}\left[d_{\rm L}<r;d_{\rm N}>d_{\rm N|L}(r)\right]}{{\rm A^L}},\nonumber\\
    =&\frac{1}{\rm A^L}\int_{h_{\rm uav}^{\rm ut}}^{\infty}\left[1-F_{d_{\rm N}}(d_{\rm N|L}(r))\right]f_{d_{\rm L}}(r) dr,
\end{align}
where the last equality follows from averaging over $d_{\rm L}$. The PDF of $\Tilde{d}_{\rm L}$ follows by differentiating with respect to $r$. The PDF of $\Tilde{d}_{\rm N}$ can be derived similarly.

\subsection{Coverage Analysis}
The coverage probability, ${\rm P^c}$, is defined as the likelihood that the received SIR across the entire service area exceeds a specified threshold, $\beta$, which is determined by the network's quality-of-service (QoS) requirements. Using the law of total probability, the coverage probability for a typical UAV-UE placed at the origin can be expressed as
\begin{equation}
    \label{eq:pctot}
    {\rm P^c} ={\rm P^c_L}\times{\rm A^L}+{\rm P^c_N}\times {\rm A^N},
\end{equation}
where ${\rm P^c_\tau}$ and ${\rm A^\tau}$, with $\tau \in \{\rm L,N\}$, are the probability of coverage of the UAV-UE given its serving ut-BS belongs to $\Phi_{\rm UT}^{\tau}$ and the association probability to that specific ut-BS tier, respectively.

\begin{theorem}
\startmycolor
The probability of coverage of the UAV-UE given that the serving ut-BS  belongs to tier  $\tau$ is expressed as 
\begin{align}
    \label{eq:pcms}
    {\rm P^c_\tau}=&\left(1- p_{\rm bm}^{\rm ut}\right)\left(1- p_{\rm bm}^{\rm uav}\right)\times\mathcal{C}_{\tau}(G_{\rm ut}G_{\rm uav})\nonumber\\
    &+\left(1- p_{\rm bm}^{\rm ut}\right) p_{\rm bm}^{\rm uav}\times\mathcal{C}_{\tau}(G_{\rm ut}g_{\rm uav})\nonumber\\
    &+p_{\rm bm}^{\rm ut}\left(1- p_{\rm bm}^{\rm uav}\right)\times\mathcal{C}_{\tau}(g_{\rm ut}G_{\rm uav})\nonumber\\
    &+p_{\rm bm}^{\rm ut} \ p_{\rm bm}^{\rm uav}\times \mathcal{C}_{\tau}(g_{\rm ut}g_{\rm uav}),
\end{align}
where $\mathcal{C}_{\tau}$ is given as
\begin{align}
\label{eq:Pcov}
\mathcal{C}_{\tau}(G_0)&=\sum_{n=0}^{m_{\tau}-1}\left(\frac{-m_{\tau}\beta}{P_{\rm ut}\left[\frac{\lambda}{4 \pi}\right]G_0 \mu_{\tau}}\right)^n\frac{1}{n!}\int_{h_{\rm uav}^{\rm ut}}^{\infty} r^{n\times \alpha_{\tau}} \nonumber\\
&\times \left[{\frac{\partial^n}{\partial s^n}}\mathscr{L}_{I|\tau}(s )\right]_{s=\left[\frac{4 \pi}{\lambda}\right]^2\frac{m_{\tau}\beta r^{\alpha_\tau}}{P_{\rm ut} \mu_{\tau} G_0}}   f_{\Tilde{d}_\tau}(r) dr
\end{align}
where $\mathscr{L}_{I|\tau}(s )=\mathscr{L}_{I^{\rm L}_{\rm UT|\tau}}(s) \times\mathscr{L}_{I^{\rm N}_{\rm UT|\tau}}(s) \times \mathscr{L}_{ I_{\rm DT|\tau}}(s)$ with $\mathscr{L}_{I^{\rm L}_{\rm UT|\tau}}(s) $, $\mathscr{L}_{I^{\rm N}_{\rm UT|\tau}}(s)$ and $\mathscr{L}_{ I_{\rm DT|\tau}}(s)$ being defined subsequently.
\end{theorem}

\textit{Proof:} See Appendix \ref{ap:pcovtau}.
\stopmycolor
\begin{table*}[t!]
\centering
\tiny
\caption{Received interference power at the UAV-UE from dt-BS used in the expression of $I_{\rm DT}$.}
\label{tab:idt}
\setlength{\tabcolsep}{2pt} 
\renewcommand{\arraystretch}{0.8} 
\begin{tabular}{l|l|l|l|l}
\hline \hline
\textbf{Condition} & \textbf{Condition on height} & \textbf{$L_1(t)$} & \textbf{$L_2(t)$} & \textbf{$L_3(t)$} \\ \hline \hline
\multirow{4}{*}{
\parbox[t]{1cm}{%
\[
\begin{aligned}
      &\theta_2^{\rm dt}\leqslant 0^{\circ}
\end{aligned}
\]
}
} &
$h_{\rm uav}<h_t$ &
$\left(\mathds{1}_{\psi_1}^{\psi_2}(t)\frac{R(t)\sqrt{G_{\rm sl}^{\rm dt}(t)}}{\sqrt{t^2+4 h_{\rm uav} h_{\rm dt}}}\right)^2$ &
$\frac{\mathds{1}_{\psi_1}^{\psi_2}(t)R(t)G_{\rm sl}^{\rm dt}(t)}{t\sqrt{t^2+4 h_{\rm uav} h_{\rm dt}}}\cos\left(\sqrt{t^2+4 h_{\rm uav} h_{\rm dt}}-t\right)$ &
\multirow{4}{*}{$\left(\frac{\sqrt{G_{\rm sl}^{\rm dt}(t)}}{t}\right)^2$} \\ \cline{2-4}
&
$h_t \leqslant h_{\rm uav} < 2 h_t$ &
$\left(\mathds{1}_{\psi_1}^{\psi_2}(t)\frac{ R(t)\sqrt{G_{\rm sl}^{\rm dt}(t)}}{\sqrt{2(t^2+4 h_{\rm uav} h_{\rm dt}})}\right)^2$ &
$\frac{\mathds{1}_{\psi_1}^{\psi_2}(t)\sqrt{2}R(t)G_{\rm sl}^{\rm dt}(t)}{t\sqrt{t^2+4 h_{\rm uav} h_{\rm dt}}}\cos\left(\sqrt{t^2+4 h_{\rm uav} h_{\rm dt}}-t\right)$ &
\\ \cline{2-4}
&
$2 h_t \leqslant h_{\rm uav} < h_{\max}$ &
$\left(\mathds{1}_{\psi_1}^{\psi_2}(t)R(t)\frac{\sqrt{G_{\rm sl}^{\rm dt}(t)-\frac{h_{\rm uav}}{h_{\max}}( G_{\rm sl}^{\rm dt}(t)-1)}}{\sqrt{2(t^2+4 h_{\rm uav} h_{\rm dt}})}\right)^2$ &
$\begin{cases}
    &\frac{\mathds{1}_{\psi_1}^{\psi_2}(t)\sqrt{G_{\rm sl}^{\rm dt}(t)}R(t)}{t}\frac{\sqrt{G_{\rm sl}^{\rm dt}(t)-\frac{h_{\rm uav}}{h_{\max}}( G_{\rm sl}^{\rm dt}(t)-1)}}{\sqrt{2(t^2+4 h_{\rm uav} h_{\rm dt})}} \\&\times \cos\left(\sqrt{t^2+4 h_{\rm uav} h_{\rm dt}}-t\right)
\end{cases}$
 &
\\ \cline{2-4}
&
$h_{\rm uav}\geqslant h_{\max}$ &
$\left(\mathds{1}_{\psi_1}^{\psi_2}(t)\frac{\sqrt{0.5}\times R(t)}{\sqrt{t^2+4 h_{\rm uav} h_{\rm dt}}}\right)^2$ &
$\frac{\mathds{1}_{\psi_1}^{\psi_2}(t)\sqrt{2}R(t)\sqrt{G_{\rm sl}^{\rm dt}(t)}}{t\sqrt{t^2+4 h_{\rm uav} h_{\rm dt}}}\cos\left(\sqrt{t^2+4 h_{\rm uav} h_{\rm dt}}-t\right)$ &
\\ \hline \hline
\multirow{4}{*}{
\parbox[t]{1cm}{%
\[
\begin{aligned}
      &\theta_2^{\rm dt}\geqslant 0^{\circ}
\end{aligned}
\]
}
} &
$h_{\rm uav}<h_t$ &
$\begin{cases}
    \left(\mathds{1}_{\psi_1}^{\psi_2}(t)\frac{R(t)\sqrt{G_{\rm sl}^{\rm dt}(t)}}{\sqrt{t^2+4 h_{\rm uav} h_{\rm dt}}}\right)^2,& t<r_2^{\rm dt}\\
    \left(\mathds{1}_{\psi_1}^{\psi_2}(t)\frac{R(t)\sqrt{G_{\rm ml}^{\rm dt}(t)}}{\sqrt{t^2+4 h_{\rm uav} h_{\rm dt}}}\right)^2,& t\geqslant r_2^{\rm dt}
\end{cases}$ &
$\begin{cases}
    \frac{\mathds{1}_{\psi_1}^{\psi_2}(t)R(t)G_{\rm sl}^{\rm dt}(t)}{t\sqrt{t^2+4 h_{\rm uav} h_{\rm dt}}}\\ \times\cos\left(\sqrt{t^2+4 h_{\rm uav} h_{\rm dt}}-t\right),& t<r_2^{\rm dt}\\
    \frac{\mathds{1}_{\psi_1}^{\psi_2}(t)R(t)G_{\rm ml}^{\rm dt}(t)}{t\sqrt{t^2+4 h_{\rm uav} h_{\rm dt}}}\\ \times\cos\left(\sqrt{t^2+4 h_{\rm uav} h_{\rm dt}}-t\right),& t\geqslant r_2^{\rm dt}
\end{cases}$ &
$\begin{cases}
    \left(\frac{G_{\rm sl}^{\rm dt}(t)}{t}\right)^2,& \\t<r_2^{\rm dt}\\
    \left(\frac{\sqrt{G_{\rm ml}^{\rm dt}(t)}}{t}\right)^2,& \\t\geqslant r_2^{\rm dt}
\end{cases}$ \\ \cline{2-4}
&
$h_t \leqslant h_{\rm uav} < 2 h_t$ &
$\begin{cases}
    \left(\mathds{1}_{\psi_1}^{\psi_2}(t)\frac{ R(t)\sqrt{G_{\rm sl}^{\rm dt}(t)}}{\sqrt{2(t^2+4 h_{\rm uav} h_{\rm dt})}}\right)^2,& t<r_2^{\rm dt}\\
    \left(\mathds{1}_{\psi_1}^{\psi_2}(t)\frac{ R(t)\sqrt{G_{\rm ml}^{\rm dt}(t)}}{\sqrt{2(t^2+4 h_{\rm uav} h_{\rm dt})}}\right)^2,& t\geqslant r_2^{\rm dt}
\end{cases}$ &
$\begin{cases}
    \frac{\mathds{1}_{\psi_1}^{\psi_2}(t)\sqrt{2}R(t)G_{\rm sl}^{\rm dt}(t)}{t\sqrt{t^2+4 h_{\rm uav} h_{\rm dt}}}\\ \times\cos\left(\sqrt{t^2+4 h_{\rm uav} h_{\rm dt}}-t\right),& t<r_2^{\rm dt}\\
    \frac{\mathds{1}_{\psi_1}^{\psi_2}(t)\sqrt{2}R(t)G_{\rm ml}^{\rm dt}(t)}{t\sqrt{t^2+4 h_{\rm uav} h_{\rm dt}}}\\ \times\cos\left(\sqrt{t^2+4 h_{\rm uav} h_{\rm dt}}-t\right),& t\geqslant r_2^{\rm dt}
\end{cases}$ &
\\ \cline{2-4}
&
$2 h_t \leqslant h_{\rm uav} < h_{\max}$ &
$\begin{cases}
    \left(\mathds{1}_{\psi_1}^{\psi_2}(t)R(t)\frac{\sqrt{G_{\rm sl}^{\rm dt}(t)-\frac{h_{\rm uav}}{h_{\max}}( G_{\rm sl}^{\rm dt}(t)-1)}}{\sqrt{2(t^2+4 h_{\rm uav} h_{\rm dt})}}\right)^2, & \\t<r_2^{\rm dt}\\
    \left(\mathds{1}_{\psi_1}^{\psi_2}(t)R(t)\frac{\sqrt{G_{\rm ml}^{\rm dt}(t)-\frac{h_{\rm uav}}{h_{\max}}( G_{\rm ml}^{\rm dt}(t)-1)}}{\sqrt{2(t^2+4 h_{\rm uav} h_{\rm dt})}}\right)^2,& \\t\geqslant r_2^{\rm dt}
\end{cases}$ &
$\begin{cases}
    \mathds{1}_{\psi_1}^{\psi_2}(t)\frac{\sqrt{G_{\rm ml}^{\rm dt}(t)}}{t}R(t)\frac{\sqrt{G_{\rm ml}^{\rm dt}(t)-\frac{h_{\rm uav}}{h_{\max}}( G_{\rm ml}^{\rm dt}(t)-1)}}{\sqrt{2(t^2+4 h_{\rm uav} h_{\rm dt})}}\\\cos\left(\sqrt{t^2+4 h_{\rm uav} h_{\rm dt}}-t\right),& t<r_2^{\rm dt}\\
    \mathds{1}_{\psi_1}^{\psi_2}(t)\frac{\sqrt{G_{\rm ml}^{\rm dt}(t)}}{t}R(t)\frac{\sqrt{G_{\rm ml}^{\rm dt}(t)-\frac{h_{\rm uav}}{h_{\max}}( G_{\rm ml}^{\rm dt}(t)-1)}}{\sqrt{2(t^2+4 h_{\rm uav} h_{\rm dt})}}\\\cos\left(\sqrt{t^2+4 h_{\rm uav} h_{\rm dt}}-t\right),& t\geqslant r_2^{\rm dt}
\end{cases}$ &
\\ \cline{2-4}
&
$h_{\rm uav}\geqslant h_{\max}$ &
$\left(\mathds{1}_{\psi_1}^{\psi_2}(t)\frac{\sqrt{0.5}\times R(t)}{\sqrt{t^2+4 h_{\rm uav} h_{\rm dt}}}\right)^2$ &
$\begin{cases}
    \frac{\mathds{1}_{\psi_1}^{\psi_2}(t)\sqrt{2}R(t)\sqrt{G_{\rm sl}^{\rm dt}(t)}}{t\sqrt{t^2+4 h_{\rm uav} h_{\rm dt}}}\cos\left(\sqrt{t^2+4 h_{\rm uav} h_{\rm dt}}-t\right),& t<r_2^{\rm dt}\\
    \frac{\mathds{1}_{\psi_1}^{\psi_2}(t)\sqrt{2}R(t)\sqrt{G_{\rm ml}^{\rm dt}(t)}}{t\sqrt{t^2+4 h_{\rm uav} h_{\rm dt}}}\cos\left(\sqrt{t^2+4 h_{\rm uav} h_{\rm dt}}-t\right),& t\geqslant r_2^{\rm dt}
\end{cases}$ &
\\ \hline \hline
\end{tabular}
\end{table*}

\startmycolor
In the following, we will derive the Laplace transform of each type of interference, accounting for the antenna patterns.
\subsubsection{ut-BSs Interference}
The typical UAV-UE is assumed to receive interference from LoS and NLoS ut-BSs. Assuming that the serving ut-BS is in tier $\tau\in\{\rm L,N\}$, the Laplace transform of the interference received from LoS non-serving ut-BSs is given by 
\begin{align}
   \mathscr{L}_{I^{\rm L}_{\rm UT|\tau}}(s) \!\!=\!\!\!\!&\prod_{\Upsilon_{\Omega}\in \Omega}\!\!\!\exp\!\Biggl(\!\!-2 \pi \lambda_{\rm ut}\mathcal{P}_{\Upsilon_{\Omega}}\!\!\!\int_{d_{{\rm L}|\tau}(r)}^{\infty} \!\!\!\!\!t\exp\left(\mathcal{K}\sqrt{t^2-(h_{\rm uav}^{\rm ut})^2}\right) \nonumber\\
    &\times \Biggl(1-\Biggl(\frac{m_{\rm L}}{m_{\rm L}+{s \left[\frac{\lambda}{4 \pi}\right]^2 P_{\rm ut}\mu_{\rm L}\Upsilon_{\Omega}t^{-\alpha_{\rm L}}}}\Biggl)^{m_{\rm L}}\Biggl)\Biggr) dt,
     \label{eq:ILOS}
\end{align}
and the Laplace transform of the interference received from NLoS non-serving ut-BSs  is given by
\begin{align}
   \mathscr{L}_{I^{\rm N}_{\rm UT|\tau}}(s) &\!=\!\!\!\!\prod_{\Upsilon_{\Omega}\in \Omega}\!\!\!\exp\!\Biggl(\!\!-\mathcal{P}_{\Upsilon_{\Omega}}\!\!\!\int_{d_{{\rm N}|\tau}(r)}^{\infty} \!\!\!\!\!t\left(1\!-\!\exp\left(\mathcal{K}\sqrt{t^2-(h_{\rm uav}^{\rm ut})^2}\right) \right)\nonumber\\
    \!\!\!\!\!\!\!\!&\!\!\!\!\!\!\!\!\!\!\!\!\times 2 \pi \lambda_{\rm ut} \Biggl(1-\Biggl(\frac{m_{\rm N}}{m_{\rm N}+{s \left[\frac{\lambda}{4 \pi}\right]^2 P_{\rm ut}\mu_{\rm N}\Upsilon_{\Omega}t^{-\alpha_{\rm N}}}}\Biggl)^{m_{\rm N}}\Biggl)\Biggr) dt,
     \label{eq:INLOS}
\end{align}
where $\mathcal{P}_{\Upsilon_{\Omega}}$  denotes the probability that the interfering ut-BS from tier $\tau$  is interfering at the UAV-UE with an antenna gain of $\Upsilon_{\Omega}$ as defined by Table \ref{tab:probint}.

\textit{Proof:} See Appendix \ref{ap:iln}.
\stopmycolor

\subsubsection{dt-BSs Interference}
The Laplace transform of the interference from dt-BSs  observed at a typical UAV-UE connected to tier $\tau$ ut-BS, due to dt-BS sidelobes and the GR of the dt-BS mainlobe, is expressed as
\startmycolor
\begin{equation}
    \label{eq:lapdt}
     \mathscr{L}_{ I_{\rm DT}|\tau}(s)\!\!=\!\!\!\! \prod_{\Psi_{\omega}\in W}\prod_{i=\{1,2,3\}} \!\!\!\!\exp\left(\!\!\int_{r}^{\infty} \!\!\!\!\frac{-2 \pi \lambda_{\rm dt}\mathcal{P}_{\Psi_{\omega}}t\ dt}{1\!+\!\left(\!\!s \!\left[\frac{\lambda}{4 \pi}\right]^2 P_{\rm dt} \Psi_{\omega}  L_i(t)\!\!\right)^{\!\!-1}} \!\!\!\right),
\end{equation}
\stopmycolor
where $L_i(t)$, $i\in\{1,2,3\}$, is specified in Table \ref{tab:idt}. \startmycolor More specifically, we can infer that $L_3(\cdot)$ denotes the interference power received from the sidelobes of the dt-BSs' radiation patterns at the UAV-UE. Whereas, $L_1(\cdot)$ is the interference power received over the reflection of the mainlobe of the dt-BSs' radiation patterns. \stopmycolor Please note that in Table \ref{tab:idt} when $L_i(t)$ is defined as a piecewise function, the integral range in the Laplace transform in eq. (\ref{eq:lapdt}) will be split into two intervals $[r, r_2^{\rm dt}]$ and $[r_2^{\rm dt},\infty]$, where $r_2^{\rm dt}$ is defined in Table \ref{tab:tab2}.

\textit{Proof:} The proof is given in Appendix \ref{ap:idt}.

Substituting the expressions  \eqref{eq:ILOS}, \eqref{eq:INLOS}, and \eqref{eq:lapdt} in (\ref{eq:Pcov}) and (\ref{eq:pcms}) yields an analytical expression for the coverage probability at the UAV-UE under the proposed architecture.  The expression incorporates two key components: 1) parameters of the dt-BS network, including downtilt and 3dB beamwidth angles, transmit power, dt-BS density, and  dt-BS height; and 2) parameters of the overlaid ut-BS network, primarily characterized by its density, transmit power, and the number of antenna elements. This expression is entirely dependent on statistical values, enabling its use for performance analysis and parameter optimization without the need for time-intensive Monte-Carlo simulations. Furthermore, optimization based on this expression does not require re-evaluation with every change in channel realization but only when the underlying channel statistics vary, which occurs on a significantly slower time scale.

Next, we present an approximation of the coverage probability in the following corollary.
\begin{corollary}
Using a tight upper bound for the CDF of the Gamma distribution \cite{alzenadcoverage}, the expression of $\mathcal{C}_{\tau}(G_0)$ in \eqref{eq:Pcov}  can be approximated as 
\begin{align}
    \label{eq:pctauapprox}
    {\hat{\mathcal{C}}_{\tau}(G_0)}\!\!\approx
        &\!\sum_{n=1}^{m_{\tau}} 
        \binom{\!m_{\tau}\!\!}{\!n}\!(\!-1\!)^{n+1}\!\!\!\!\int_{h_{\rm uav}^{\rm ut}}^{\infty}\!\!\!\!\!\!\mathscr{L}_{\!I|\tau}\!\!\left(\!\!\frac{n m_{\tau}(m_{\tau}!)^{\frac{-1}{m_{\tau}}}\beta r^{\alpha_{\tau}}}{\left[\frac{\lambda}{4 \pi}\right]^2 P_{\rm ut}\mu_{\tau}G_{0} }\!\!\right)\!\!f_{\!\Tilde{d}_\tau}(\!r\!)dr
\end{align}\end{corollary}
\textit{Proof:} See \cite[Appendix F]{alzenadcoverage}.\\
This simplification reduces computational complexity and enhances numerical stability, making the coverage probability expression more practical for implementation. The resulting expression enables faster and more reliable analysis by replacing complex derivatives with an efficient and relatively accurate approximation.
\stopmycolor

\begin{table}[t!]
\centering
\footnotesize
\renewcommand{\arraystretch}{0.7} 
\setlength{\tabcolsep}{3pt} 
\caption{Simulation Parameters}
\begin{tabular}{l l | l l}
\hline
\textbf{Parameter} & \textbf{Value} & \textbf{Parameter} & \textbf{Value} \\ \hline
$(\theta_{\rm dt},h_{\rm dt})$ & $(-15^{\circ},25$ m) & $\theta^{\rm dt}_{\rm 3dB}$ & $10^{\circ}$ \\ 
\textcolor{black}{$(m_{\rm L},m_{\rm N})$} & \textcolor{black}{(3,1)} & $(\alpha_{\rm L},\alpha_{\rm N})$ & $(2, 3.5)$ \\ 
$(\mu_{\rm L},\mu_{\rm N})$ & $(-1\,\text{dB}, -20\,\text{dB})$ \cite{alhouranilet} & $f_c$ & 2 GHz \\ 
$(\lambda_{\rm dt},\lambda_{\rm ut})$ & (10 dt-BS/km$^2$, 2 ut-BS/km$^2$) & $(P_{\rm dt},P_{\rm ut})$ & (20 W, 10 W) \\ 
$(N_{\rm ut},N_{\rm uav})$ & (16, 1) & $\beta$ & $-5\,\text{dB}$ \\ 
Ground & Medium dry ground & \textcolor{black}{$(\nu,\kappa,\xi)$} & \textcolor{black}{(0.3,500,15) \cite{alhouranilet}} \\ \hline
\end{tabular}
\label{tab:simparam}
\vspace{10pt}
\end{table}

It is worth noting that the proposed stochastic geometry  framework is intended for SIR-based coverage analysis. The objective is to derive a tractable network-level coverage expression for a typical UAV-UE located anywhere in this network at a given height $h_{\rm uav}$, while accounting for key effects such as LoS/NLoS propagation, radiation patterns, and interference from non-serving ut-BSs, sidelobes of the dt-BSs and GR of the mainlobe of the dt-BSs.  We also incorporated beam misalignment caused by UAV mobility  in the system model through eq. \eqref{eq:pbm} and the subsequent analysis. While there are also other important aspects that can affect the performance  at the UAV-UE such as  orientation dynamics, Doppler effects, handovers, and synchronization impairments, incorporating each of these effects would require a different analytical framework with  temporal mobility and channel evolution modeling. These can be important directions in which this analysis can be extended further.  Moreover, channel state information (CSI) acquisition and the associated overhead are also important aspects to consider in future works that perform the net rate analysis for this architecture. This overhead, however, does not impact the  SIR-based coverage metric considered in this work \cite{andrews,venugopal2016device,alzenad,aquino2015}.
\section{Optimization of ut-BS Network Parameters}
Designating some dt-BSs additionally as ut-BSs by installing an uptilted antenna array in addition to the downtilted antenna array already present can result in harmful interference at the GUE. This interference typically comes from the sidelobes in the ut-BS's radiation pattern defined in Sec. II-C.  Hence, the design of this hybrid network infrastructure for aerial coverage should account for how much degradation it can cause to cellular ground coverage.
\begin{figure}
    \centering
    \includegraphics[width=\linewidth]{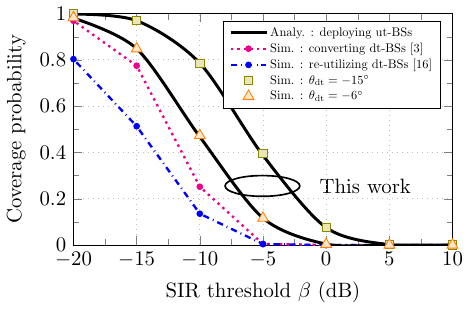}
    \caption{Probability of coverage at UAV-UE versus $\beta$.}
    \label{fig:qos}
    \vspace{15pt}
\end{figure}
In this section, we formulate an optimization problem to design various parameters of the ut-BS network to maximize the probability of coverage at the UAV-UE under GUE performance constraints. The probability of coverage at the GUE in the absence of ut-BS network is given in \cite[eq.(28)]{aquino2015}, and it is denoted in this work as $\rm P^{c, gue}$. We define the GUE coverage in the presence of ut-BS network as $\rm P^{c, gue}_{\rm ut}$, which can be easily obtained by incorporating the ut-BS interference in  $\rm P^{c, gue}$ through its Laplace transform that can be derived similarly to the results in Sec. III as 
\begin{align}
    \label{eq:LIutGUE}
    \mathscr{L}_{\rm UT}^{\rm gue}(r)=\exp\left(\!\!\int_{r}^{\infty} \!\!\!\!\frac{-2 \pi \lambda_{\rm ut}t}{1+\frac{P_{\rm dt}G_{\rm tot}^{\rm dt}(r) r^{-\alpha_{\rm N}}}{P_{\rm ut}\beta  g_{\rm ut}t^{-\alpha}}} dt\right),
\end{align}
 where $r$ in eq. \eqref{eq:LIutGUE} denotes the distance between the GUE and the closest interfering ut-BS.

 We aim to maximize the UAV-UE coverage probability while ensuring that the performance degradation at the GUE remains within the maximum allowable limit $\varsigma \leq 1$. The problem is formulated as
\begin{align}
\label{eq:optimization}
  {\rm (P1)} \hspace{.2in}  \lambda_{\rm ut}^*,h_{\rm uav}^*,P_{\rm ut}^*&=\underset{\lambda_{\rm ut},h_{\rm uav},P_{\rm ut}}{\rm arg\!\!\!
    \!\quad max} \quad{\rm P^{c}}\nonumber\\
    \text{s.t} & \quad 
    \frac{\rm P^{c,gue}_{\rm ut}}{\rm P^{c,gue}} \geqslant \varsigma .
\end{align}
 We define $ \rm P^{c,*} $ as the maximum coverage probability achievable at the UAV-UE by selecting the optimal values for ut-BS network parameters, i.e., ut-BS density  $ \lambda_{\rm ut}^*$, the UAV-UE height  $ h_{\rm uav}^* $, and the ut-BS transmit power $P_{\rm ut}^*$. The values are selected such that the decrease in coverage probability at the GUE is no less than a preset threshold $\varsigma$.  \startmycolor Finding a global solution to (P1) using the expression derived in Theorem 1 is extremely challenging due to the complex interdependence of the variables, as well as the presence of numerical integrals and products within the expressions. To address this, we employ a linear search strategy, which is both tractable and efficient, and widely used in similar optimization problems \cite{aquino2015,kalamkar2022beam}. Specifically, we optimize each variable independently through linear search while keeping the other two variables fixed, leveraging ${\text{MATHEMATICA}}^{\tiny\textcircled{c}}$ for computation. This approach proves effective in determining the optimal values, which are presented and discussed in detail in the numerical results section. We would also like to highlight that the solution of the optimization problem, although numerically computed, only depends on network parameters and channel statistics. Thus, it only needs to be re-computed once the channel statistics or network change. The simplicity and computational efficiency of linear search make it particularly well-suited for solving such problems. \stopmycolor The network operators can utilize the solution of this problem to plan the deployment of ut-BSs on a portion of terrestrial infrastructure in a given coverage area.

\section{Numerical Results}
 The simulation parameters are summarized in Table \ref{tab:simparam}. \textcolor{black}{Throughout this section, we represent the Monte-Carlo simulated results for the coverage probability  with ``markers" and the analytical expression with ``lines".}

\subsection{Merits of collocating ut-BSs for aerial coverage and impact of dt-BS tilt}
Fig. \ref{fig:qos} illustrates the probability of coverage at UAV-UE  versus different QoS requirements represented by $\beta$. The derived analytical expression in eq. \eqref{eq:pctot} computed using the expressions in eqs \eqref{eq:Pcov}-(\ref{eq:lapdt}) is in good agreement with the Monte-Carlo simulated curve.  \textcolor{black}{Moreover, in Fig. \ref{fig:qos}, we simulate the approach in \cite{chen32023}, where 20\% of the dt-BSs were converted to BSs equipped with uptilted antennas. The magenta curve represents this setup, using parameters $(\theta_t,\phi_{\rm 3dB})=(15^\circ,10^\circ)$ \cite{chen32023}, where $\theta_t$ is the uptilt angle and $\phi_{\rm 3dB}$ is the HPBW angle. As illustrated in the figure, collocating directional uptilted antennas on 20\% of the dt-BSs outperforms 1) converting 20\% of the dt-BSs to uptilted BSs serving UAV-UE \cite{chen32023} or 2) relying on the dt-BSs sidelobes for aerial coverage \cite{Cherif2025}. \textcolor{black}{This improvement is mainly attributed to our use of highly directional uptilted antennas, specifically, a UPA rather than the ULA considered in \cite{chen32023}, which concentrates radiated power toward the UAV-UE, thereby strengthening the desired signal and enhancing the resulting SIR.} While our collocation strategy requires the deployment of additional  antenna arrays on a portion of the network, as opposed to simply converting some dt-BSs to ut-BSs electrically, the collocation strategy not only enhances aerial coverage  but also ensure that the GUE performance is not impacted significantly. This is in contrast to the conversion approach in \cite{chen32023}, which will result in the loss in performance of GUE since a smaller number of dt-BSs will now serve it. This will be shown in the subsequent simulation results. Moreover, the UAV-UE coverage deteriorates with a smaller downtilt angle $\theta_{\rm dt}$ of the dt-BSs. This is mainly because a smaller downtilt angle results in a small $\psi_1$ (See Fig. \ref{fig:geometry_sys}), which is the angle of reference between the ground and the region on the sky where the ground reflected mainlobe interference has the most impact. This will result in a more significant impact of the ground reflected mainlobe interference when $\theta_{\rm dt}=-6^o$, even when the UAV-UE has a small elevation angle with respect to the BS.}

\subsection{Energy efficiency of  the proposed architecture vs baselines}
\begin{figure}[t!]
    \centering
    \includegraphics[width=\linewidth]{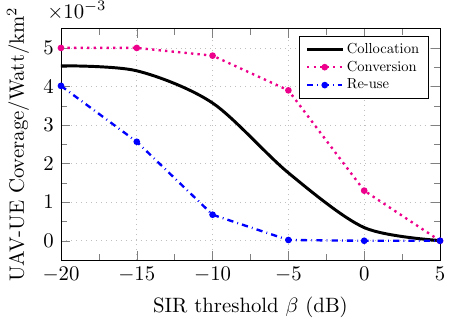}
    \caption{UAV-UE coverage per Watt per km$^2$.}
    \label{fig:cov_eff}
    \vspace{15pt}
\end{figure}

In Fig. \ref{fig:cov_eff}, we plot the UAV-UE coverage per Watt per km$^2$ versus the UAV-UE SIR threshold. This metric is defined as the ratio between the UAV-UE coverage probability and the total radiated power (in Watts) within a ($1~\mathrm{km}^2$) area. For example, under the dt-BS reuse baseline in \cite{Cherif2025} and the antenna-conversion baseline in \cite{chen32023}, the total radiated power equals ($\lambda_{\rm dt}P_{\rm dt}$
). In contrast, under the proposed partial collocation architecture, where additional uptilted antenna arrays are deployed, the total radiated power becomes ($\lambda_{\rm dt}P_{\rm dt}+\lambda_{\rm ut}P_{\rm ut}$).

The results indicate that the conversion-based architecture in \cite{chen32023} attains the largest UAV-UE coverage per transmitted Watt per km$^2$ among the considered schemes, and in particular outperforms the proposed collocation approach. This advantage stems from the fact that collocation requires extra transmitted power due to the newly installed uptilted antenna arrays, whereas the conversion approach enhances aerial coverage by reconfiguring a fraction of existing downtilted BSs into uptilted BSs without increasing the total number of antenna arrays.

Nevertheless, from the GUE perspective, the collocation architecture preserves both the original power budget for dt-BSs and the infrastructure dedicated to GUE service. By contrast, the conversion strategy reallocates a subset of downtilted BSs to aerial coverage, which reduces the resources available for GUE performance. The clear advantage of collocation strategy in terms of preserving GUE performance will become explicit in the subsequent figures.
\stopmycolor

\subsection{Impact of $h_{\rm uav}$ and environment on the UAV-UE coverage}
\begin{figure}[t!]
    \centering
    \includegraphics[width=\linewidth]{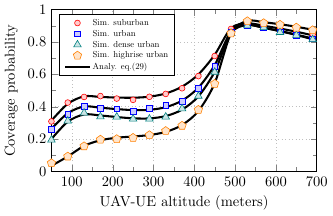}
    \caption{Probability of coverage for different environments.}
    \label{fig:env}
    \vspace{10pt}
\end{figure}

Fig. \ref{fig:env} shows the evolution of the coverage probability as a function of the UAV-UE height and the terrestrial environment nature. We see that the analytical expression in eq. \eqref{eq:pctot} matches the Monte-Carlo simulations for any UAV-UE height or environment.  The environment (e.g., suburban, urban, dense urban) mainly determines the likelihood of the association of the UAV-UE with a LoS or NLoS ut-BS.  The denser the environment, the smaller is the UAV-UE coverage performance, since the likelihood of the LoS path between the serving ut-BS and UAV-UE decreases. This is in contrast to what is observed in the literature, where we see that the UAV-UE, when being served by the dt-BS, achieves better coverage probability in denser environments. This is because it benefits from smaller interference in denser environments, thanks to more obstruction resulting in interference originating from NLoS paths instead of LoS paths \cite{cherifdownlink,alzenadcoverage}. 

Interestingly, we observe that the coverage probability at the UAV-UE increases with the height despite the increase in path loss. This is explained mainly by the decrease in the GR-induced interference (from the mainlobes of the radiation patterns of the dt-BSs) received at the UAV-UE at elevated heights (which is corroborated in eq. (\ref{eq:gain_ref})). Note that at $h_{\max}=500$ meters height,  the GR gain reduces to only $0.5$ dB, and we see that the coverage probability at the UAV-UE now decreases with height, regardless of the environment, owing to the increase in the path loss.  
 \begin{figure}[t!]
     \centering
     \includegraphics[width=\linewidth]{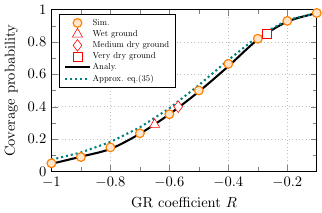}
     \caption{Probability of coverage versus the GR coefficient ($h_{\rm uav}=300$ m.).}
     \label{fig:grcoeff}
     \vspace{10pt}
 \end{figure}


\subsection{Impact of the GR coefficient }
The impact of GR depends on the strength of the  paths propagating along the mainlobe of the dt-BS radiation pattern towards the ground, the UAV-UE height, and the GR coefficient $R$ that in turns depends on the antenna polarization and the permittivity of the ground \cite[Table 7.6.8-1]{tr138901}. We investigate the coverage probability at the UAV-UE as a function of the GR coefficient in Fig. \ref{fig:grcoeff}. It is worth highlighting that all the simulated coefficients may not have been derived from real-world scenarios. However, we consider a wide range of coefficients to understand how the GR impacts the UAV-UE coverage probability. We also mark on the figure the real-world GR coefficients for different types of ground  (medium dry, wet, and very dry) \cite[Table 7.6.8-1]{tr138901}.  As shown in Fig. \ref{fig:grcoeff}, increasing the GR coefficient $R$ improves the UAV-UE coverage probability. This trend is consistent with the physical interpretation of the reflection model in \cite{tr138901} (see [eqs. (7.6-38)-(7.6-42)]). In particular, larger values of $R$ correspond to ground conditions with lower relative permittivity (e.g., very dry terrain). Under such conditions, the electromagnetic field associated with the mainlobe of the radiation pattern of the interfering dt-BS is less efficiently reflected at the air-ground interface, and instead tends to penetrate further into the ground, where it is partially absorbed and/or scattered. As a result, the reflected radiation pattern component contributing to the dt-BS induced interference is weakened, which reduces the aggregate interference term $I_{\rm DT}$ experienced by the UAV-UE at elevated heights, and therefore yields a higher coverage probability. \stopmycolor 

\textcolor{black}{Lastly, the dotted curve indicates that the approximation presented in Corollary 1 closely matches the exact analytical expression of the coverage probability, while avoiding the need for extensive numerical computations. \textcolor{black}{Thus, Corollary 1 can be used as an alternative to accurately evaluate the coverage probability.}}

\begin{figure}[t!]
    \centering
    \includegraphics[width=\linewidth]{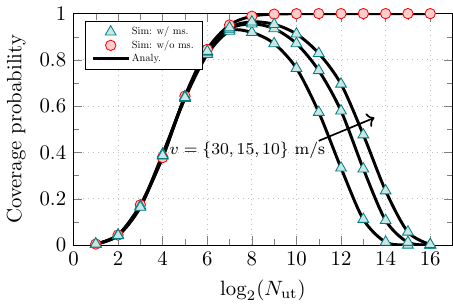}
    \caption{Probability of coverage at UAV-UE versus $\log_2(N_{\rm ut})$ with and without misalignment (ms.) with $t_{\rm SSB}=20$ ms.}
    \label{fig:ms}
    \vspace{10pt}
\end{figure}
\begin{figure}[t!]
    \centering
    \includegraphics[width=\linewidth]{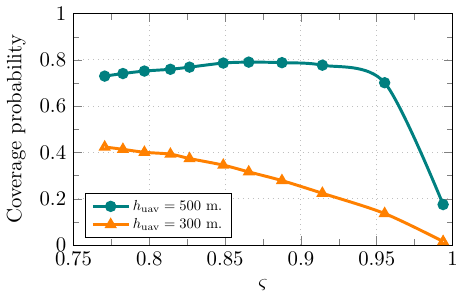}
    \caption{Probability of coverage at UAV-UE versus $\varsigma$ in urban environment.}
    \label{fig:varsigma}
    \vspace{10pt}
\end{figure}

\subsection{Beam misalignment effect}

We analyze the impact of $ \log_2(N_{\rm ut}) $ on the occurrence of beam misalignment and its effect on UAV-UE performance in Fig. \ref{fig:ms}. The results indicate that for a relatively small number of antennas, beam misalignment is highly unlikely and therefore has a negligible effect on the UAV-UE coverage probability. In contrast, when the array size increases to 256 antennas, i.e., $\log_2(N_{\rm ut})=8$, beam misalignment becomes more pronounced and leads to a noticeable reduction in coverage probability compared to the ideal case without misalignment. Moreover, this degradation increases with the UAV-UE velocity, since higher speeds raise the probability of misalignment between the serving ut-BS and the typical UAV-UE.

While beam misalignment does not pose a problem at lower frequencies, where we work with antenna arrays of up to 64 antennas, it will affect the coverage probability even under the collocation scheme at higher frequencies, utilizing ultra-massive MIMO arrays with more than 256 antennas. Therefore, enabling ubiquitous and reliable aerial coverage at these frequencies will not only require better network architectures (such as the collocation scheme) but also better beam alignment and tracking algorithms for aerial users.
\stopmycolor

\subsection{Optimization of the ut-BS network parameters  }

Fig.~\ref{fig:varsigma} illustrates the coverage probability of the UAV-UE as a function of the GUE degradation threshold 
$\varsigma$, where the constraint in (P1) is enforced by appropriately adjusting the density of ut-BSs, $\lambda_{\rm ut}$ for given $P_{\rm ut}$ and $h_{\rm uav}$. As 
$\varsigma$ increases, the requirement on the GUE coverage becomes more stringent (i.e., less degradation is allowed), which restricts the number of the overlaid ut-BSs in the network. At an altitude of 300 m, this leads to a monotonic decrease in the UAV-UE coverage probability with $\varsigma$, since the reduction in ut-BS density primarily weakens the desired signal at the UAV-UE. At  500 m, however, the UAV-UE coverage exhibits a non-monotonic behavior and is maximized around $\varsigma=0.87$, corresponding to an allowable GUE degradation of about 
13 \%. The latter is explained by noting that while increasing $\varsigma$ leads to a sparser ut-BS deployment (smaller $\lambda_{\rm ut}$) that  decreases the desired signal strength, it also decreases interference that now comes from a smaller number of non-serving ut-BSs. The overall impact of  these two effects on UAV-UE coverage depends on the UAV altitude.  At $300$ m, where the GR interference from dt-BSs is dominant, tightening the constraint (larger $\varsigma$), and hence reducing $\lambda_{\rm ut}$, decreases the desired signal more than it mitigates interference from non-serving ut-BSs, which results in a lower UAV-UE coverage probability. By contrast, at $500$ m, where GR interference becomes negligible and the dominant impairment is the interference from non-serving ut-BSs, increasing $\varsigma$ initially improves coverage because the reduction in ut-BS density decreases the interference more rapidly than the desired signal. However, beyond a certain value of $\varsigma$, the serving-signal loss caused by further reducing $\lambda_{\rm ut}$ becomes dominant, leading to a decrease in UAV-UE coverage probability.

\stopmycolor

\begin{figure}[t!]
    \centering
    \includegraphics[width=\linewidth]{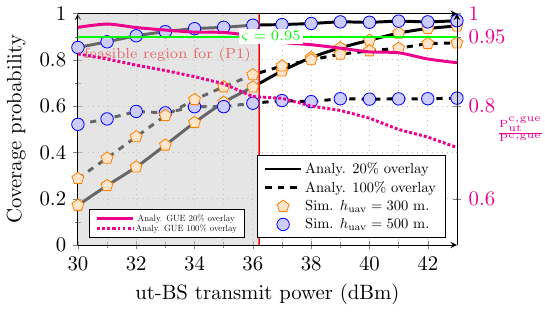}
    \caption{Probability of coverage versus the ut-BS transmit power, $P_{\rm ut}$, in suburban environment.}
    \label{fig:put}
    \vspace{10pt}
\end{figure}

Fig.~\ref{fig:put} shows the coverage probability of the UAV-UE on left y-axis and the ratio of coverage at GUE with and without ut-BSs on right y-axis as a function of the ut-BS transmit power for different overlay percentages. An overlay percentage of $20$ \% means that only $20$ \% of the existing dt-BSs are additionally equipped with uptilted antenna arrays, whereas 
$100$ \% overlay corresponds to ut-arrays collocated at all dt-BSs. As expected, the UAV-UE coverage probability increases with the ut-BS transmit power for a fixed overlay percentage, due to the stronger desired signal. Moreover, for a $20$ \%
 overlay, the UAV-UE achieves better coverage at an altitude of 500 m than at 300 m, since the impact of GR is reduced at $h_{\max} = 500$ meters. In contrast, when the overlay percentage is $100$ \%, the UAV-UE at 300 m outperforms that at
500 m. This is because even though the downlink terrestrial interference is more severe at 
300 m, the gain in desired signal strength at the lower UAV-UE altitude with higher ut-BS density more than compensates for the increased interference.

Interestingly, as the ut-BSs transmit power exceeds 38 dBm, the coverage probability at the UAV-UE at the height of 300 m becomes higher when the overlay percentage is 20 \% compared to that at an overlay percentage of 100 \%. This counterintuitive result is because as transmit power increases, the interference from a larger number of non-serving ut-BSs when the overlay percentage is higher causes the coverage probability at the UAV-UE to reduce. This shows the inherent trade-offs between desired signal and various forms of interference at different UAV-UE heights and overlay percentages in determining UAV-UE performance.

On the other hand, the GUE coverage probability decreases as the ut-BS transmit power grows, with a more pronounced degradation for the 
100 \% overlay case, reflecting the stronger interference injected into the terrestrial network by the ut-BSs sidelobes. In Fig.~\ref{fig:put}, we also highlight the feasible region of problem (P1) for $\varsigma=0.95$, i.e., when no more than a 5 \% loss in GUE coverage is allowed with the deployment of ut-BSs. Within this region, the trade-offs among ut-BS power, overlay percentage, and UAV-UE altitude become explicit. In particular, we observe that choosing an overlay percentage of about 
20 \%, a UAV-UE altitude of approximately 
500 m, and a ut-BS transmit power of 36 dBm yields a near-optimal operating point that maximizes the UAV-UE coverage while satisfying the constraint in (P1) on GUE performance.
\stopmycolor

\section{Conclusion}
In this study, we adapted the terrestrial network architecture to support cellular coverage for UAV-UE by equipping some existing BS towers, referred to as dt-BSs, with  additional antenna arrays of uptilted antennas. These BSs were additionally referred to as ut-BSs. Using stochastic geometry tools, we analyzed the coverage probability of a typical UAV-UE served by these ut-BSs under a LoS/NLoS probabilistic link model between each ut-BS and UAV-UE. The analysis accounted for accurate radiation patterns at the dt-BSs and ut-BSs, \textcolor{black}{sidelobe interference caused by dt-BSs} and non-serving ut-BSs, as well as GR-induced interference from the mainlobe of the dt-BSs' radiation patterns. The derived expression was used to formulate an optimization problem that finds the optimal parameters of the ut-BS network, such as overlay percentage, height of UAV-UE, and transmit power of ut-BSs, to maximize the coverage probability at the UAV-UE while constraining the impact on the GUE to be within a threshold. Our simulation results
demonstrated that overlaying a small portion of the dt-BS network with uptilted antennas can significantly improve aerial coverage, thanks to the strong direct link that ut-BSs
provide. Furthermore, we showed that this deployment results in a small reduction in the coverage performance of the GUE due to the interference caused by the sidelobes of the ut-BSs, whereas the interference from the sidelobes of the dt-BSs has only a small impact on the performance of the UAV-UE. However, the ground reflected mainlobe can introduce significant interference  at lower UAV heights, which is why operating  UAV-UE at relatively higher altitudes improves the coverage probability despite the larger path loss in the ut-BS UAV-UE link. For the considered simulation framework, the optimal ut-BS deployment parameters that maximize the coverage probability at the UAV-UE without degrading the GUE performance by more than 5\%  turned out to be $(\lambda_{\rm ut}^{*},h_{\rm uav}^{*},\ P_{\rm ut}^*)=(2 \text{ ut-BSs/km}^2, 500 \text{ meters},36 \text{ dBm})$. These findings offer valuable guidelines for telecommunication operators interested in overlaying a portion of the existing dt-BS network with uptilted antenna arrays to enhance aerial coverage in the future, as UAVs find more use cases and applications.

\appendices

\renewcommand{\thesection}{\Alph{section}}
 \section{}
 The derivation of distance and angle in \eqref{eq:r1r2} and \eqref{eq:refangle} is provided below.
\label{ap:geometry}
As shown in Fig. \ref{fig:geometry_sys}, $d$ is the 2D distance separating the BS radio tower from the UAV-UE.  We can easily express $d$ as a function of the altitude difference between the dt-BS and the UAV-UE as well as the 3D distance between them, by applying the Pythagorean theorem as 
\begin{equation}
    \label{eqap:d2}
    d^2=l_j^2-(h_{\rm uav}-h_{\rm dt})^2.
\end{equation}
Invoking the Pythagorean theorem once more, we can express the total reflected distance, $x=r_{1,j}+r_{2,j}$, as
\begin{equation}
    \label{eqap:x2}
    x^2=d^2+(h_{\rm uav}+h_{\rm dt})^2.
\end{equation}
Then, substituting eq. (\ref{eqap:d2}) into eq. (\ref{eqap:x2}) along with some mathematical simplification yields eq. (\ref{eq:r1r2}). For the reflection angle $\epsilon_j$, we can easily express it using the definition of $\sin$ function as 
\begin{equation}
    \label{eqap:epsilon}
    \sin \epsilon_j=\frac{h_{\rm uav}+h_{\rm dt}}{x}.
\end{equation}
Substituting the expression of $x$  in eq. (\ref{eqap:epsilon}) completes the proof.

\section{}
\label{ap:frl}
Since $d_{\rm L}$ is the closest 3D distance between the UAV-UE and the LoS ut-BS, we define the 2D distance as $z_{\rm L}=\sqrt{d_{\rm L}^2-(h_{\rm uav}^{\rm ut})^2}$. By invoking the null probability of PPP \cite{andrews}, we have
\begin{eqnarray}
F_{d_{\rm L}}(r)&=&1-\mathbb{P}\left(d_{\rm L}\geqslant r\right)\nonumber\\
&=&1-\exp\left(-2\pi\lambda_{\rm ut}\int_{h_{\rm uav}^{\rm ut}}^{r} t \mathcal{P}_{\rm L}(t)dt\right).\nonumber\\
\end{eqnarray}
Substituting $r=\sqrt{z^2+(h_{\rm uav}^{\rm ut})^2}$ and using  integral by parts to solve the identity $\int_{h_{\rm uav}^{\rm ut}}^r t \exp(\mathcal{K} \sqrt{t^2-(h_{\rm uav}^{\rm ut})^2})dt = \frac{1+(-1+\mathcal{K} \sqrt{r^2-(h_{\rm uav}^{\rm ut})^2})\exp[{\mathcal{K} \sqrt{r^2-(h_{\rm uav}^{\rm ut})^2}}]}{\mathcal{K}^2}$ completes the CDF derivation. The PDF is easily obtained by differentiating with respect to $r$.

\section{}
\label{ap:pcovtau}
\startmycolor
Following the discrete random variable $G_0$ defined in Table \ref{tab:ms} that represents \textcolor{black}{combined antenna gain of the serving ut-BS UAV-UE link}, we can express the probability of coverage of the UAV-UE conditioned on being served by ut-BS from tier $\tau$ as
\begin{equation}
    \label{apeq:pctau1}
    {\rm P^c_\tau}=\sum_{G_0} \mathcal{P}_{G_0} \mathcal{C}_{\tau}(G_0),
\end{equation}
where $\mathcal{C}_{\tau}(G_0)$ can be expressed using  eq. (\ref{eq:SIR}) as
\stopmycolor
\startmycolor
\begin{align}
    \label{apeq:pctau}
    &\mathcal{C}_{\tau}(G_0)=\mathbb{E}_{q^{\tau}_{{\rm ut},x^*},\Tilde{d}_\tau,I}\left[\mathbb{P}\left({\frac{P_{\rm ut} \left[\frac{\lambda}{4 \pi}\right]^2 \mu_{\tau} G_0 q^{\tau}_{{\rm ut},x^*} r_{\tau}^{-\alpha_{\tau}}}{I} \geqslant \beta}\right)\right]\nonumber\\
      &=  \mathbb{E}_{\Tilde{d}_\tau}\left[\mathbb{E}_{I}\left[\frac{\Gamma\left(m_{\tau},\left[\frac{4 \pi}{\lambda}\right]^2\frac{m_{\tau}\beta r_{\tau}^{\alpha_\tau} I}{P_{\rm ut} \mu_{\tau} G_0}\right)}{\Gamma(m_{\tau})}\right]\right]\nonumber\\
        &= \underset{n\in [0,m_{\tau}-1]}{\sum}\left(\frac{m_{\tau}\beta}{P_{\rm ut}\left[\frac{\lambda}{4 \pi}\right]^2G_0 \mu_{\tau}}\right)^n\frac{1}{n!}   \nonumber\\
    &\times \mathbb{E}_{\Tilde{d}_\tau}\left[ r_{\tau}^{n\times \alpha_{\tau}}\mathbb{E}_{I}\left[I^n\exp\left(- \frac{m_{\tau} \beta r_{\tau}^{\alpha_{\tau}}}{P_{\rm ut}\left[\frac{\lambda}{4 \pi}\right]^2G_0 \mu_{\tau}}I\right)\right]\right],
\end{align}
where the second equality in eq. (\ref{apeq:pctau}) is obtained thanks to the independence between the PPP $\Phi_{\rm UT}^{\tau}$, the interference $I$, and  the Gamma-distributed channel gain. Moreover, applying the CCDF of the Gamma distribution and assuming an integer $m_{\tau}$ parameter yields eq. \eqref{apeq:pctau}. Finally, applying the identity $\mathbb{E}_I\left[I^n \exp(-sI)\right]=(-1)^n {\frac{\partial^n}{\partial s^n}}\mathscr{L}_{I}(s )$, averaging over $\Tilde{d}_\tau(r)$, and invoking the independence between the dt-BSs interference and the non-serving LoS and NLoS ut-BS interference result in eq. \eqref{eq:Pcov}.
\section{}
\label{ap:iln}
The Laplace transform of the LoS interference from non-serving LoS ut-BSs set $\Phi_{\rm UT}^{\rm L}$, given that the serving ut-BS belongs to tier $\Phi_{\rm UT}^{\tau}$ where $\tau \in \{\rm L, N\}$, is expressed as
\begin{align}
   &\mathscr{L}_{I_{\rm UT|\tau}^{\rm L}}\!(s)\!\!=\!\!\mathbb{E}_{\Phi_{\rm UT}^{\rm L}}\!\!\!\left[ \!\underset{x_i \in \Phi_{\rm UT}^{\rm L} \backslash x^*}{\prod} \!\!\!\!\!\! \mathbb{E}_{ q_{{\rm ut},i}^{\rm L}}\! \left[\! \exp\left(\!\!-sP_{\rm ut}\left[\!\!\frac{\lambda}{4 \pi}\!\!\right]^2\!\!\Upsilon_{\Omega} q_{{\rm ut},i}^{\rm L} r_{i,\rm L}^{-\alpha_{\rm L}}\!\!\right)\!\!\right]\!\!\right].
   \label{apeq:Iut}
\end{align}
Since the channel gain is independent for all $i\in \Phi_{\rm UT}^{\rm L}$ and using the moment generating function (MGF) of the Gamma distribution, we simplify the inner expectation in eq. (\ref{apeq:Iut}), to obtain
\begin{align}
   &\mathscr{L}_{I_{\rm UT|\tau}^{\rm L}}\!\!(\!s\!)\!\!=\!\mathbb{E}_{\Phi_{\rm UT}^{\rm L}}\!\!\!\left[\!\underset{x_i \in \Phi_{\rm UT}^{\rm L} \backslash x^*}{\prod} \!\!\!\!\left(\frac{m_{\rm L}}{m_{\rm L}+s P_{\rm ut}\left[\frac{\lambda}{4 \pi}\right]^2\Upsilon_{\Omega} \ t^{-\alpha_{\rm L}}}\!\!\right)^{\!\!\!m_{\rm L}} \!\right]\!\!.
   \label{apeq:Iut1}
\end{align}
By invoking the definitions of the generating functional of a PPP \cite{cherifdownlink} and the discrete random variable $\Upsilon_{\Omega}$ with probability distribution $\mathcal{P}_{\Upsilon_{\Omega}}$ (defined in Table \ref{tab:probint}), we obtain
\textcolor{black}{
\begin{align}
   &\mathscr{L}_{I_{\rm UT|\tau}^{\rm L}}(r)\!\!=\!\!\exp \!\! \left(\!\sum_{\Upsilon_{\Omega} \in \Omega}\!\! -2\pi \lambda_{\rm ut}\mathcal{P}_{\Upsilon_{\Omega}}\!\!\displaystyle\int_{d_{{\rm L}|\tau}(r)}^{\infty}\!\!\!t\exp\!\!\left(\!\mathcal{K}\sqrt{t^2\!-\!(\!h_{\rm uav}^{\rm ut}\!)^2}\right)\!\!\!\right)\nonumber\\
    &\times \Biggl(1-\Biggl(\frac{m_{\rm L}}{m_{\rm L}+ s P_{\rm ut}\left[\frac{\lambda}{4 \pi}\right]^2\Upsilon_{\Omega} \ t^{-\alpha_{\rm L}} }\Biggl)^{m_{\rm L}}\Biggl)dt\Biggr).
   \label{apeq:Iut2}
\end{align}}

$\mathscr{L}_{I^{\rm N}_{\rm UT|\tau}}(r)$ can be derived similarly.
\stopmycolor
\section{}
 \label{ap:idt}
We start by simplifying the received power from the GR of the mainlobe of the dt-BSs radiation pattern, given in eq. (\ref{eq:IGR}) with $\alpha({h_{\rm uav}})=2$ (refer to eq. \eqref{eq:alphah}), and using the identity $|a+b\times e^{ic}|^2=a^2+b^2+2\times a\times b\cos c$ as  
\begin{align}
    \label{eqap:idt}
    &\mathscr{L}_{I_{\rm DT|\tau}}(s)
    = \mathbb{E}_{q_{{\rm dt},j}, l_j} \Biggl[
    \exp\Biggl( -s\underset{x_j \in \Phi_{\rm DT} \backslash x^*}{\sum} P_{\rm dt}\left[\frac{\lambda}{4 \pi}\right]^2q_{{\rm dt},j}\Psi_{\omega} \nonumber\\[1mm]
    &\quad \times \Biggl[ \left(\frac{\sqrt{G_{\rm tot}^{\rm dt}(l_j)}}{l_j}\right)^2
    + \left( \mathds{1}_{\psi_1}^{\psi_2}(l_j)\,\frac{0.5\, R(l_j)\, \sqrt{G_{\rm ref}^{\rm dt}(l_j)}}{\sqrt{l_j^2+4\, h_{\rm uav}\, h_{\rm dt}}} \right)^2\nonumber\\
    &\quad+ 2 \times \frac{\sqrt{G_{\rm tot}^{\rm dt}(l_j)}}{l_j}\times \mathds{1}_{\psi_1}^{\psi_2}(l_j)\,\frac{0.5\, R(l_j)\, \sqrt{G_{\rm ref}^{\rm dt}(l_j)}}{\sqrt{l_j^2+4\, h_{\rm uav}\, h_{\rm dt}}} \nonumber\\
    &\quad\times \cos\left(\sqrt{l_j^2+4 h_{\rm uav} h_{\rm dt}}-l_j\right)\Biggr]
    \Biggr) \Biggr].
\end{align}

We then leverage two key notions: 1) the independence of the PPP, which is captured through the distance $l_j$, and the small-scale fading represented by $q_{{\rm dt},j}$, and 2) the moment-generating function of the exponential distribution.  Next, by applying the Probability Generating Functional (PGFL) \cite{andrews}, and averaging over the PDF of the distance $l_j$, as defined in eq. (\ref{eq:fdt}), we derive the result presented in eq. (\ref{eq:lapdt}) and in Table \ref{tab:idt}.\stopmycolor

\bibliographystyle{IEEEtran}  
\bibliography{references,Bibliography}

\end{document}